\documentclass[sigconf,natbib=true,screen]{acmart}
\usepackage{xspace,balance,tabularx,multirow}
\usepackage{flushend}
\usepackage{tikz}
\usepackage{pgfplots}
\pgfplotsset{compat=1.16}
\usetikzlibrary{patterns}
\usepackage{subfig}
\usepackage[ruled, vlined, linesnumbered]{algorithm2e}
\usepackage{xcolor}
\usepackage{colortbl}
\usepackage{bbold}
\SetKwComment{Comment}{$\triangleright$\ }{}
\usepackage{enumitem}
\usepackage{tablefootnote}
\usepackage{upgreek,textgreek}
\usepackage{pifont}%
\usepackage[noabbrev]{cleveref}
\usepackage{titlecaps}
\usepackage{lipsum}

\pgfplotsset{every tick label/.append style={font=\tiny}}

\newlength{\starsize}
\newlength{\starspread}
\tikzset{starsize/.code={\setlength{\starsize}{#1}},
         starspread/.code={\setlength{\starspread}{#1}}}
\tikzset{starsize=1mm,
         starspread=3mm}
\pgfdeclarepatternformonly[\starspread,\starsize]%
  {my fivepointed stars}%
  {\pgfpointorigin}%
  {\pgfqpoint{\starspread}{\starspread}}%
  {\pgfqpoint{\starspread}{\starspread}}%
  {%
   \pgftransformshift{\pgfqpoint{\starsize}{\starsize}}
   \pgfpathmoveto{\pgfqpointpolar{18}{\starsize}}
   \pgfpathlineto{\pgfqpointpolar{162}{\starsize}}
   \pgfpathlineto{\pgfqpointpolar{306}{\starsize}}
   \pgfpathlineto{\pgfqpointpolar{90}{\starsize}}
   \pgfpathlineto{\pgfqpointpolar{234}{\starsize}}
   \pgfpathclose%
   \pgfusepath{fill}
  }

\makeatletter
\newcommand*\bigcdot{\mathpalette\bigcdot@{.5}}
\newcommand*\bigcdot@[2]{\mathbin{\vcenter{\hbox{\scalebox{#2}{$\m@th#1\bullet$}}}}}
\makeatother

\newcommand{\stitle}[1]{\vspace*{0.5em}\noindent{\bf #1.\/}}

\newcommand{\U}{\mathcal{U}\xspace}
\newcommand{\V}{\mathcal{C}\xspace}
\newcommand{\G}{\mathcal{G}\xspace}
\newcommand{\C}{\mathcal{B}\xspace}
\newcommand{\M}{\mathcal{Y}\xspace}

\newcommand{\N}{\mathcal{N}\xspace}
\newcommand{\EDG}{\mathcal{E}\xspace}

\newcommand{\AM}{\mathbf{A}\xspace}
\newcommand{\NAM}{\mathbf{\tilde{A}}\xspace}
\newcommand{\DM}{\mathbf{D}\xspace}
\newcommand{\IM}{\mathbf{I}\xspace}
\newcommand{\SM}{\mathbf{S}\xspace}
\newcommand{\MM}{\mathbf{M}\xspace}
\newcommand{\EM}{\mathbf{E}\xspace}
\newcommand{\PM}{\mathbf{P}\xspace}
\newcommand{\CM}{\mathbf{C}\xspace}
\newcommand{\YM}{\mathbf{Y}\xspace}

\newcommand{\LM}{\mathbf{L}\xspace}
\newcommand{\UM}{\mathbf{U}\xspace}

\newcommand{\GM}{\mathbf{G}\xspace}
\newcommand{\BM}{\mathbf{X}\xspace}

\newcommand{\dvec}{\mathbf{d}\xspace}
\newcommand{\dtvec}{\boldsymbol{\delta}\xspace}
\newcommand{\svec}{\boldsymbol{\sigma}\xspace}

\newcommand{\PiM}{\mathbf{G}\xspace}

\newcommand{\DeltaM}{\boldsymbol{\Phi}\xspace}

\newcommand{\xvec}{\mathbf{x}\xspace}

\newcommand{\algo}{\texttt{CASO}\xspace}

\newcommand{\eat}[1]{}

\SetKwInOut{Parameter}{Parameters}

\newenvironment{customlegend}[1][]{%
    \begingroup
    \csname pgfplots@init@cleared@structures\endcsname
    \pgfplotsset{#1}%
}{%
    \csname pgfplots@createlegend\endcsname
    \endgroup
}%

\def\addlegendimage{\csname pgfplots@addlegendimage\endcsname}

\makeatletter
\newcommand\footnoteref[1]{\protected@xdef\@thefnmark{\ref{#1}}\@footnotemark}
\makeatother

\let\oldnl\nl%
\newcommand{\nonl}{\renewcommand{\nl}{\let\nl\oldnl}}%

\SetKwComment{Comment}{/* }{ */}

\makeatletter 
\g@addto@macro{\@algocf@init}{\SetKwInOut{Parameter}{Parameters}} 
\makeatother

\definecolor{myred}{HTML}{fd7f6f}
\definecolor{myred_new}{HTML}{D8D8D8}
\definecolor{myred_new2}{HTML}{D7191C}
\definecolor{myblue}{HTML}{7eb0d5}
\definecolor{mygreen}{HTML}{b2e061}
\definecolor{mypurple}{HTML}{bd7ebe}
\definecolor{myorange}{HTML}{ffb55a}
\definecolor{myyellow}{HTML}{ffee65}
\definecolor{mypurple2}{HTML}{beb9db}
\definecolor{mypink}{HTML}{fdcce5}
\definecolor{mycyan}{HTML}{8bd3c7}

\definecolor{myblue2}{HTML}{115f9a}
\definecolor{myred2}{HTML}{c23728}

\definecolor{bblue}{HTML}{1f77b4}
\definecolor{rred}{HTML}{C0504D}
\definecolor{ggreen}{HTML}{9BBB59}

\definecolor{NSCcol1}{HTML}{1f77b4}
\definecolor{NSCcol2}{HTML}{aec7e8}
\definecolor{NSCcol3}{HTML}{ff7f0e}
\definecolor{NSCcol4}{HTML}{ffbb78}
\definecolor{NSCcol5}{HTML}{98df8a}
\definecolor{NSCcol6}{HTML}{2ca02c}

\AtBeginDocument{%
  \providecommand\BibTeX{{%
    \normalfont B\kern-0.5em{\scshape i\kern-0.25em b}\kern-0.8em\TeX}}}

\setcopyright{acmcopyright}
\copyrightyear{2018}
\acmYear{2018}
\acmDOI{XXXXXXX.XXXXXXX}

\acmISBN{978-1-4503-XXXX-X/18/06}

\acmSubmissionID{8875}

\begin{document}

\title{Community-Aware Social Community Recommendation}
\subtitle{Technical Report}

\author{Runhao Jiang}
\affiliation{%
  \institution{Hong Kong Baptist University}
  \country{Hong Kong SAR, China}
}
\email{csrhjiang@comp.hkbu.edu.hk}

\author{Renchi Yang}
\authornote{Corresponding Author}
\affiliation{%
  \institution{Hong Kong Baptist University}
  \country{Hong Kong SAR, China}
}
\email{renchi@hkbu.edu.hk}
\orcid{0000-0002-7284-3096}

\author{Wenqing Lin}
\affiliation{%
  \institution{JD.com}
  \country{China}
}
\email{linwenqing.8@jd.com}

\renewcommand{\shortauthors}{Trovato and Tobin, et al.}

\begin{abstract}
Social recommendation, which seeks to leverage social ties among users to alleviate the sparsity issue of user-item interactions, has emerged as a popular technique for elevating personalized services in recommender systems.
Despite being effective, existing social recommendation models are mainly devised for recommending regular items such as blogs, images, and products, and largely fail for community recommendations due to overlooking the unique characteristics of communities. Distinctly, communities are constituted by individuals, who present high dynamicity and relate to rich structural patterns in social networks.
To our knowledge, limited research has been devoted to comprehensively exploiting this information for recommending communities.

To bridge this gap, this paper presents \algo{}, a novel and effective model specially designed for social community recommendation. Under the hood, \algo{} harnesses three carefully-crafted encoders for user embedding, wherein two of them extract community-related global and local structures from the social network via {\em social modularity maximization} and {\em social closeness aggregation}, while the third one captures user preferences using collaborative filtering with observed user-community affiliations.
To further eliminate feature redundancy therein,
we introduce a mutual exclusion between social and collaborative signals.
Finally, \algo{} includes a community detection loss in the model optimization, thereby producing community-aware embeddings for communities.
Our extensive experiments evaluating \algo{} against nine strong baselines on six real-world social networks demonstrate its consistent and remarkable superiority over the state of the art in terms of community recommendation performance.
\end{abstract}



\maketitle

\section{Introduction}
In social media platforms, users who share similar interests tend to cluster together~\cite {anagnostopoulos2008influence}, forming groups, clubs, circles, cohorts, forums, etc. 
Such communities find extensive applications, including marketing and advertising~\cite{bakhthemmat2021communities}, user tagging~\cite{tan2021multiple}, recommender systems~\cite{gasparetti2021community,zhang2023constrained}, and many other personalized services~\cite{wu2006research}.
Although a cornucopia of community detection techniques~\cite{girvan2002community, clauset2004finding, blondel2008fast, jin2019graph, jia2019communitygan} has been invented in the past two decades, they mainly focus on grouping users from a global perspective without personalization and hardly deal with the ever-increasing volume and dynamics of social networks and communities.
Moreover, real social networks often comprise a multitude of implicit or explicit communities (e.g., user-created groups or clubs) and observed user-community memberships, which cannot be used by this methodology.
In recent years, {\em social community recommendation}~\cite{chen2008combinational_ccf,zhang2023constrained} task has been framed to generate personalized community recommendations to social users, thereby promoting community growth for better online services~\cite{backstrom2006group} in social networks, e.g., LinkedIn~\cite{sharma2013pairwise_PLSI} and Tencent Games~\cite{zhang2023constrained}.

Early works~\cite{chen2008combinational_ccf,chen2009collaborative,sharma2013pairwise_PLSI,velichety2020finding,wang2016recommending} towards community recommendation are mainly based on {\em collaborative filtering} (CF), which predicts a user's preferences on communities by leveraging the observed community memberships of similar users.
However, the performance of these methods is less than satisfactory due to the high sparsity of the user-community interaction data. As a partial remedy, several attempts~\cite{chen2008combinational_ccf,lu2023ksgan,sharma2013pairwise_PLSI,velichety2020finding} have been made to capitalize on auxiliary information for enhanced recommendation.
For instance, \cite{chen2008combinational_ccf} and \cite{velichety2020finding} additionally consider community descriptions or attributes, while \cite{sharma2013pairwise_PLSI} extracts negative samples from impression logs and \cite{lu2023ksgan} profiles the information of scholarly communities from scholarly knowledge graphs, respectively.

The emergence and remarkable potency of {\em graph neural networks} (GNNs) \cite{kipf2016semi} have enabled us to tap into the rich semantics underlying the sheer amount of social relations in social networks for recommendation, which inspire a series of graph-based social recommendation models~\cite{fan2019GraphRec,wu2019diffnet,wu2020diffnet++,yu2021MHCN,yu2021SEPT,fan2019graph,sharma2024survey}.
Basically, these models learn user latent factors from the social network and user-item interaction graph through the {\em neighborhood aggregation} scheme widely adopted in GNNs. 
This paradigm is well suited for characterizing influence diffusion in the social context~\cite{fan2019graph,tao2022revisiting}, due to its alignment with the {\em social homophily}~\cite{mcpherson2001birds_homophily} and {\em social influence} theories~\cite{marsden1993network}, which state that socially-connected users are likely to have similar preferences and behaviors of a user can be influenced by his/her friends.
The majority of these approaches either explore various neighborhood aggregation schemes \cite{wu2019diffnet,yu2021MHCN,wu2020diffnet++,fan2019GraphRec,yu2021SEPT} or focus on denoising social relations for a better alignment with user preferences~\cite{quan2023robust,sun2023denoising,yang2024GBSR,jiang2024challenging}.

Although these methods obtain improved performance in recommending regular items, they are suboptimal for community recommendations due to the intrinsic differences between communities and other items. 
Particularly, contrary to regular items, communities are highly dynamic and should be characterized by their respective members (i.e., social users). In addition, each community presents informative connectivity patterns of inter- and intra-community users on the social network. 
Recently, \citet{zhang2023constrained} extend the foregoing models to community recommendation by aggregating features from distant neighborhoods based on {\em personalized PageRank}~\cite{jeh2003scaling} and intra-community neighbors on the social network extended with community nodes.
Nevertheless, this approach focuses on the constraint setting where each user can join at most one community, and also fails to capture the unique characteristics of social communities adequately.
To our knowledge, none of the existing works conduct an in-depth study of exploiting the community-aware structural patterns for social community recommendation over social networks.

To bridge this gap, this paper proposes  \algo{} (\underline{C}ommunity-\underline{A}ware \underline{S}ocial c\underline{O}mmunity recommendation), a novel and effective framework specially catered for social community recommendation.
The design motivation of \algo{} is grounded on three community-related structural patterns empirically observed in real social networks and user-community affiliations, which are quantified by our proposed measures: {\em average connectivity}, {\em common neighbors}, and {\em common communities}.
Thereon, we develop four key components in \algo{}, including three disentangled encoders for users: {\em social modularity maximization} (SMM), {\em social closeness aggregation} (SCA), {\em user-based collaborative encoding} (UCE), and one module for {\em feature mutual exclusion} (FME).
More specifically, SMM seeks to optimize the {\em modularity}~\cite{newman2004finding} of user embeddings over the social network, thereby incorporating the global social patterns of communities (i.e., inter- and intra-community connectivity) into the user features, whereas SCA captures the local structures in social networks by leveraging the {\em neighborhood-based social closeness}~\cite{lu2011link} for user feature aggregation.
On the other hand, UCE injects the collaborative signals into user embeddings using a collaborative encoding of observed user-community memberships with an ameliorated preference similarity metric.
To maximize the complementary efficacy of social and collaborative features, FME refines user embeddings based on optimizing the empirical HSIC~\cite{gretton2005measuring_HSIC} of them for feature redundancy reduction and higher expressive power.
Further, \algo{} includes a classic Kullback-Leibler (KL) divergence loss for community detection in model optimization, so as to inject the subordinate or inclusion relationships between users and communities for the generation of communities.

Our extensive experiments comparing \algo{} against nine baselines on six real-world social networks exhibit that our proposed \algo{} consistently and considerably outperforms all competitors in terms of recall and NDCG. In particular, on the well-known {\em Youtube} dataset~\cite{yang2012defining-largedata}, \algo{} is able to achieve a margin of $2.54\%$ in Recall@3 and $5.38\%$ in NDCG@3 over the state of the art.

\section{Related Work}

\subsection{Community Recommendation}
Early works mainly employed CF to mine the information of user-community memberships for community recommendations.  
\texttt{CCF}~\cite{chen2008combinational_ccf} combines information on both community-user and community-description co-occurrences, thereby enabling more effective personalized recommendations.
Pairwise \texttt{PLSI}~\cite{sharma2013pairwise_PLSI} employs a probabilistic latent semantic indexing model in the pairwise learning approach to mitigate the adverse effects caused by users' preferences.
\cite{wang2016recommending} captures the temporally extended nature of group engagement by modeling the implicit user-community affinity.
On top of that, \cite{velichety2020finding,lu2023ksgan} have achieved enhanced performance by leveraging auxiliary information such as community attributes and knowledge graphs.
Recently, \citet{zhang2023constrained} propose to extend social networks with community nodes and user-community links and apply GNNs over the extended graph for recommendation.
However, this work only considers the constrained problem where each user can only join at most one community, and also fails to exploit the intrinsic social structures of communities.

\subsection{Social Recommendation}
Inspired by the {\em social homophily theory} \cite{mcpherson2001birds_homophily} and {\em social influence theory} \cite{marsden1993network}, social recommendation methods aim to combine social relations and user preferences for better recommendations.
Initial attempts, such as SoRec \cite{ma2008sorec} and SocialMF \cite{jamali2010matrix}, leverage social relations in a simplistic way, mainly relying on co-factorization methods \cite{konstas2009social, ma2008sorec} and regularization-based strategies \cite{jamali2010matrix, jiang2014scalable, ma2011recommender}.

Over the past few years, due to the remarkable success of GNNs, graph-based social recommendations have been the focus of extensive research \cite{fan2019GraphRec, wu2019diffnet,wu2020diffnet++, yu2021MHCN, yang2023hyperbolic,yang2024GBSR}.
For instance, \texttt{GraphRec}~\cite{fan2019GraphRec} constructs user embeddings by aggregating features from two perspectives.
\texttt{DiffNet}~\cite{wu2019diffnet} models the recursive dynamic social diffusion in social networks, and \texttt{DiffNet++} \cite{wu2020diffnet++} utilizes a hierarchical attention mechanism to jointly incorporate social and interaction influences.
Instead of learning in Euclidean space, several studies have ventured into hyperbolic learning methodologies specifically tailored for graph-based social recommendation systems \cite{wang2021hypersorec, yang2023hyperbolic}.
\cite{yu2021MHCN, yu2021SEPT} have combined hypergraph structures to extract high-order information from social networks, aided by self-supervised learning techniques.
Recent studies~\cite{yang2024GBSR, quan2023robust, sun2023denoising} have revealed the low preference-aware homophily in social networks and are committed to developing denoising mechanisms for robust recommendations.
State-of-the-art models are mainly designed for regular items in recommender systems, which disregard the unique traits of social communities, and thus, fail to adequately leverage pertinent social and collaborative signals.

\subsection{Community Detection}
Community detection seeks to partition the users in social networks into multiple groups (a.k.a. communities, clusters).
Traditional approaches harness graph partitioning and hierarchical clustering, with seminal algorithms like the Girvan-Newman method \cite{girvan2002community} iteratively removing high-betweenness edges to uncover communities. Modularity optimization, notably advanced by Newman \cite{newman2004finding}, became a cornerstone, with greedy techniques \cite{clauset2004finding} and the Louvain algorithm \cite{blondel2008fast} enhancing scalability. To find overlapping communities, clique-based methods like the Clique Percolation Method~\cite{palla2005uncovering} and label propagation algorithms \cite{raghavan2007near_LPA} emerged, addressing the limitations of disjoint partitions. 

With the rise of deep learning, recent approaches resort to and adapt various deep learning models to encode nodes in social graphs into feature vectors for clustering, such as topology-recovering convolutional neural networks \cite{xin2017deep, cai2020edge}, {\em graph convolutional networks} \cite{jin2019graph}, and {\em graph attention networks}~\cite{fu2020magnn}.
In addition, CommunityGAN \cite{jia2019communitygan} utilizes generative adversarial networks to further enhance discriminative power for overlapping communities through synthetic sample generation, while autoencoder-based models~\cite{yang2016modularity} and \cite{choong2018learning} focus on reconstructing network embeddings.
Moreover, deep {\em nonnegative matrix factorization} (NMF) extends traditional NMF methods with hierarchical factorizations to capture multi-level community patterns~\cite{ye2018deep,huang2021community}, and {\em deep sparse filtering} addresses data sparsity through sparsity-aware embedding techniques \cite{xie2018community}. 
Community detection has been extensively studied in the literature, and detailed surveys can be found in \cite{bedi2016community,su2022comprehensive,jin2021survey,yang2021effective,yang2024efficient}.

\section{Preliminaries}

Let $\G=(\U,\EDG)$ be a social network (a.k.a. social graph), wherein $\U=\{u_1,u_2,\ldots,u_{|\U|}\}$ is a set of users and $\EDG \subseteq \U\times \U$ is a set of friendships between users. For each friendship $(u_i,u_j)\in \EDG$, we say $u_i$ and $u_j$ are neighbors to each other, and we use $\N(u_i)$ to denote the set of neighbors of $u_i$. The degree of user $u_i$ is symbolized by $\dvec_i=|\N_{\G}(u_i)|$ where $\dvec$ is a length-$|\U|$ column vector. We represent the adjacency matrix of $\G$ by $\AM\in \mathbb{R}^{|\U|\times |\U|}$ and by $\DM=\texttt{diag}(\dvec)$ the diagonal degree matrix.
Accordingly, matrices $\PM=\DM^{-1}\AM$ and $\NAM=\DM^{-1/2}\AM\DM^{-1/2}$ stand for the transition matrix and normalized adjacency matrix of $\G$, respectively.

 Let $\C=(\U\cup \V,\M)$ be a {\em community membership network} (CMN), in which $\U$ is a set of users and $\V=\{c_1,c_2,\ldots,c_{|\V|}\}$ represent a set of communities (a.k.a. groups, clubs). We denote by $\M\subseteq \U\times \V$ the community memberships of users and by $\YM\in \mathbb{R}^{|\U|\times |\V|}$ the bi-adjacency matrix of $\C$, where $\YM_{i,k}=1$ if user $u_i$ belongs to community $c_k$, i.e., $(u_i,c_k)\in \M$, and 0 otherwise.
The set of communities containing $u_i$ can be symbolized by $\N_{\C}(u_i)$ and the set of members in community $c_k$ is represented by $\N_{\C}(c_k)$. 
We use $\dtvec_i=|\N_{\C}(u_i)|$ and $\svec_k=|\N_{\C}(c_k)|$ to symbolize the degree of user $u_i$ and the size of community $c_k$ in CMN $\C$, respectively, and $\dtvec$ and $\svec$ are two column vectors.

\stitle{AC, ACN, and ACC}
Given a set $\V=\{c_1,c_2,\ldots,c_{|\V|}\}$ of communities, we define the {\em average connectivity} (AC) for intra-community users and inter-community users over $\G$ as follows:
\begin{small}
\begin{equation*}
\begin{gathered}
\textnormal{AC}_{\textnormal{intra}}=\frac{\underset{c_k \in \V}{\sum}\underset{\ u_i,u_j\in c_k}{\sum}{\mathbb{1}_{(u_i,u_j)\in \EDG}}}{\underset{c_k \in \V}{\sum}{\svec_k\cdot (\svec_k-1)}}, \textnormal{AC}_{\textnormal{inter}}=\frac{\underset{c_k, c_\ell \in \V}{\sum}\underset{\ u_i\in c_k, u_j \in c_\ell}{\sum}{\mathbb{1}_{(u_i,u_j)\in \EDG}}}{\underset{{c_k, c_\ell \in \V}}{\sum}{\svec_k\cdot \svec_\ell}},
\end{gathered}
\end{equation*}
\end{small}
where $\mathbb{1}_{(u_i,u_j)\in \EDG}=1$ is an indicator function, whose value is $1$ if $(u_i,u_j)\in \EDG$ and $0$ otherwise. Intuitively, $\textnormal{AC}_{\textnormal{intra}}$ (resp. $\textnormal{AC}_{\textnormal{inter}}$) calculates the averaged fraction of user pairs within a community (resp. across different communities) that are friends.

Analogously, we define {\em average common neighbors} (ACN) for intra-community users and inter-community users: 
\begin{small}
\begin{equation*}
\begin{gathered}
\textnormal{ACN}_{\textnormal{intra}}=\frac{\underset{c_k \in \V}{\sum}\underset{\ u_i,u_j\in c_k}{\sum}{|\N_\G(u_i)\cap \N_\G(u_j)|}}{\underset{c_k \in \V}{\sum}{\svec_k\cdot (\svec_k-1)}},\\
\textnormal{ACN}_{\textnormal{inter}}=\frac{\underset{c_k, c_\ell \in \V}{\sum}\underset{\ u_i\in c_k, u_j \in c_\ell}{\sum}{|\N_\G(u_i)\cap \N_\G(u_j)|}}{\underset{{c_k, c_\ell \in \V}}{\sum}{\svec_k\cdot \svec_\ell}}, 
\end{gathered}
\end{equation*}
\end{small}
which quantify their average numbers of common neighbors in $\G$, respectively. In the same vein, the {\em average common communities} (ACC) in $\C$ for intra-community users and inter-community users can be formulated by
\begin{small}
\begin{equation*}
\begin{gathered}
\textnormal{ACC}_{\textnormal{intra}}=\frac{\underset{c_k \in \V}{\sum}\underset{\ u_i,u_j\in c_k}{\sum}{|\N_\C(u_i)\cap \N_\C(u_j)|-1}}{\underset{c_k \in \V}{\sum}{\svec_k\cdot (\svec_k-1)}},\\
\textnormal{ACC}_{\textnormal{inter}}=\frac{\underset{c_k, c_\ell \in \V}{\sum}\underset{\ u_i\in c_k, u_j \in c_\ell}{\sum}{|\N_\C(u_i)\cap \N_\C(u_j)|}}{\underset{{c_k, c_\ell \in \V}}{\sum}{\svec_k\cdot \svec_\ell}}.
\end{gathered}
\end{equation*}
\end{small}
Note that $|\N_\C(u_i)\cap \N_\C(u_j)|-1$ in $\textnormal{ACC}_{\textnormal{intra}}$ is to exclude the currently known community $c_k$ where both $u_i$ and $u_j$ coexist. 

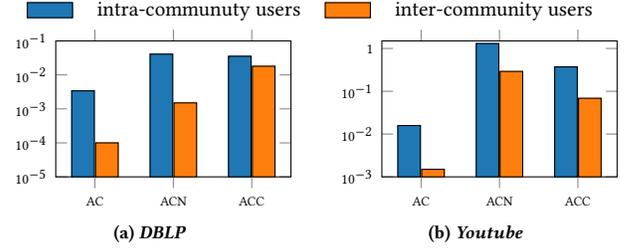
\begin{figure}[!t]
  \centering
\begin{tikzpicture}
\begin{customlegend}[
        legend entries={intra-communuty users,inter-community users},
        legend columns=2,
        area legend,
        legend style={at={(0.45,1.15)},anchor=north,draw=none,font=\small,column sep=0.25cm}]
        \addlegendimage{preaction={fill, NSCcol1}} 
        \addlegendimage{preaction={fill, NSCcol3}}    
    \end{customlegend}
\end{tikzpicture}
\\[-\lineskip]
\vspace{-4mm}
\subfloat[{\em DBLP}]{
\begin{tikzpicture}
    \begin{axis}[
        height=\columnwidth/2.5,
        width=\columnwidth/1.8,
        xtick=\empty,
        ybar=0.5pt,
        bar width=0.3cm,
        enlarge x limits=true,
        symbolic x coords={AC,ACN,ACC},
        xtick = data,
        enlarge x limits=0.25,
        ymode=log,
        log origin y=infty,
        log basis y={10},
        ymin=1e-5,
        ymax=0.1,
        ytick={1e-5,1e-4,1e-3,1e-2,1e-1},
        yticklabels={$10^{-5}$,$10^{-4}$,$10^{-3}$,$10^{-2}$,$10^{-1}$},
    ]
    \addplot[preaction={fill, NSCcol1}] coordinates {(AC, 0.0034) (ACN,0.0409) (ACC,0.0354)};
    \addplot[preaction={fill, NSCcol3}] coordinates {(AC,0.0001) (ACN,0.0015) (ACC,0.0180)};
    \end{axis}
\end{tikzpicture}
}\hspace{4mm}%
\subfloat[{\em Youtube}]{
\begin{tikzpicture}
    \begin{axis}[
        height=\columnwidth/2.5,
        width=\columnwidth/1.8,
        xtick=\empty,
        ybar=0.5pt,
        bar width=0.3cm,
        enlarge x limits=true,
        symbolic x coords={AC,ACN,ACC},
        xtick = data,
        enlarge x limits=0.25,
        ymode=log,
        log origin y=infty,
        log basis y={10},
        ymin=1e-3,
        ymax=1.5,
        ytick={1e-3,1e-2,1e-1,1},
        yticklabels={$10^{-3}$,$10^{-2}$,$10^{-1}$,$1$},
    ]
    \addplot[preaction={fill, NSCcol1}] coordinates {(AC, 0.0158) (ACN,1.3039) (ACC,0.3726)};
    \addplot[preaction={fill, NSCcol3}] coordinates {(AC,0.0015) (ACN,0.2907) (ACC,0.0685)};
    \end{axis}
\end{tikzpicture}
}%
\vspace{-2ex}
\caption{The AC, ACN, and ACC for intra- and inter-community nodes on {\em DBLP} and {\em Youtube}.}
\label{fig:motivation}
\end{figure}

\stitle{Modularity} The {\em modularity} introduced by ~\citet{newman2004finding} aims at quantifying the goodness (i.e., internal and external connectivity) of a particular division of a network. Formally, given social network $\G$ and CMN $\C$, the modularity $Q(\AM)$ is defined by
\begin{small}
\begin{equation*}
Q(\AM) = \sum_{(u_i,u_j)\in \EDG}{\left( \AM_{i,j} - \frac{{\dvec_i\cdot \dvec_j}}{|\EDG|}\right) \cdot \phi(u_i,u_j)},
\end{equation*}
\end{small}
where $\phi(u_i,u_j)=1$ if $\exists c_k\in \V$ s.t. $\BM_{i,k}=\BM_{j,k}=1$, i.e., users $u_i, u_j$ are in the same community, and otherwise $\phi(u_i,u_j)=0$. Particularly, a larger modularity value indicates a better partitioning of the social network.

In addition, the standard modularity can be easily extended to its normalized version as follows so as to achieve a balanced weighting of friendships of users with numerous and scarce friends.
\begin{small}
\begin{equation}\label{eq:norm_modularity}
Q(\NAM) = \sum_{(u_i,u_j)\in \EDG}{\left( \NAM_{i,j} - \frac{\sqrt{\dvec_i}\cdot \sqrt{\dvec_j}}{|\EDG|}\right) \cdot \phi(u_i,u_j)}.
\end{equation}
\end{small}

\stitle{Problem Statement}
Given a social network $\G$ and an incomplete CMN $\C$ as input, the goal of social community recommendation is to predict the missing user-community memberships $\overline{\M}$ in $\M$.

\section{Methodology}

\subsection{Motivation and Framework Overview}\label{sec:overview}
Before delving into the details of our proposed \algo{} model, we first delineate the design motivation and an overview of \algo{}.

\stitle{Preliminary Empirical Study}
In real social networks, communities are groups of individuals who share common preferences and/or form intimate connections. Intuitively, users therein tend to present the following community-related structural patterns:
\begin{itemize}[leftmargin=*]
\item {\bf Global Social Patterns}: Users in the same community are internally connected via friendships, and meanwhile, are externally disconnected.
\item {\bf Local Social Patterns}: Users with more common neighbors (i.e., friends) are more likely to join the same communities. 
\item {\bf Collaborative Patterns}: Users with similar community memberships are more likely to join the same communities.
\end{itemize}

To validate these patterns, we report the AC, ACN, and ACC for intra- and inter-community users on real-world social networks {\em DBLP}~\cite{yang2012defining-largedata} and {\em Youtube}~\cite{yang2012defining-largedata}, respectively, in Fig.~\ref{fig:motivation}. It can be observed that compared to $\textnormal{AC}_{\textnormal{inter}}$, $\textnormal{ACN}_{\textnormal{inter}}$, and $\textnormal{ACC}_{\textnormal{inter}}$, which evaluate the AC, ACN, and ACC of users across different communities, $\textnormal{AC}_{\textnormal{intra}}$, $\textnormal{ACN}_{\textnormal{intra}}$, and $\textnormal{ACC}_{\textnormal{intra}}$ are conspicuously higher on both datasets. The results indicate that users within the same community have more friendship connections, share more common friends, and are more likely to join the same communities, which are consistent with the aforementioned structural patterns.

\begin{figure}[!t]
\centering
    \includegraphics[width=0.9\columnwidth]{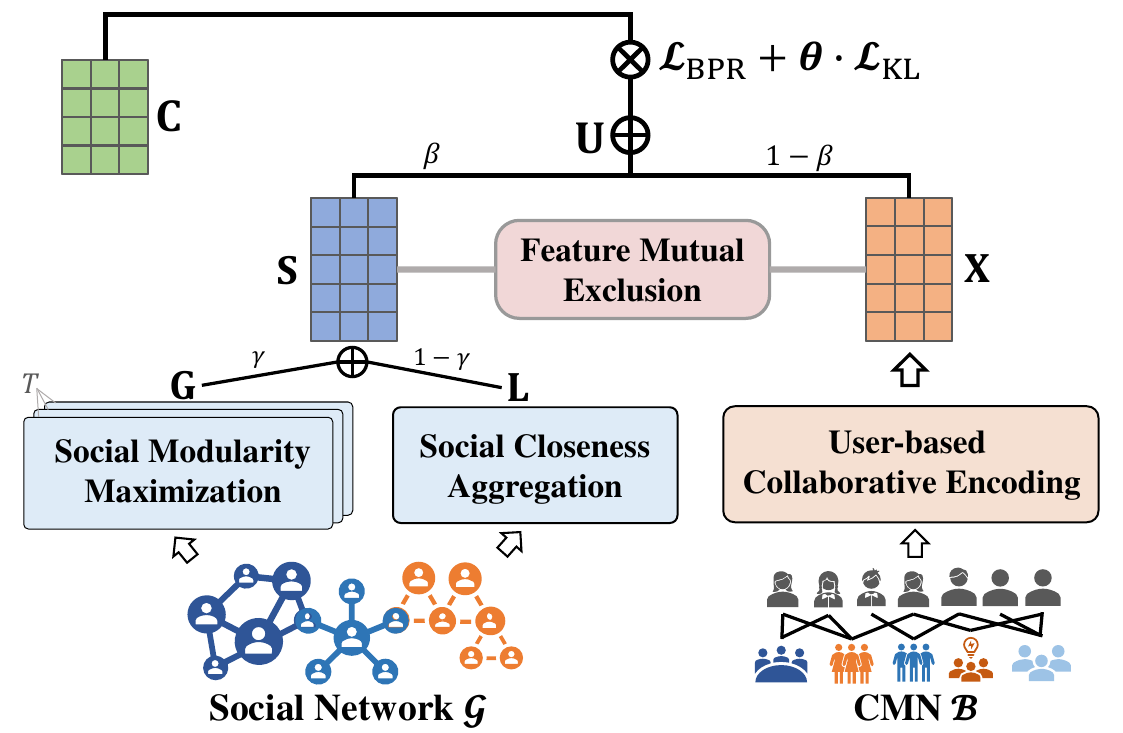}
    \vspace{-2ex}
    \caption{The overall framework of \algo{}.}
    \label{fig:overview}
    \vspace{-3ex}
\end{figure}

\stitle{Synoptic Overview} Motivated by such observations, our framework \algo{} seeks to exploit the global and local social structures underlying social network $\G$ and collaborative signals in CMN $\C$ that pertain to communities, such that they can effectively complement each other for better community recommendation.
As summarized in Fig.~\ref{fig:overview}, \algo{} follows the popular {\em two-tower architecture}~\cite{covington2016deep} in recommender systems, where users and communities are separated into two individual encoders to reduce online computational complexity.
Notice that \algo{} mainly focuses on encoding the foregoing structural features for users, since communities are radically distinct from the items in most recommender systems, which are dynamic and characterized by users.
We instead introduce a community detection loss $\mathcal{L}_{\textnormal{KL}}$ in Section~\ref{sec:training} for the refinement of community features.

More specifically, given the randomly initialized user embeddings $\UM^{\circ}\in \mathbb{R}^{|\U|\times d}$, \algo{} harnesses two encoders based on {\em social modularity maximization} (SMM) and {\em social closeness aggregation} (SCA) to construct $\GM$ and $\LM$, which incorporate the global and local social patterns underlying $\G$, respectively. They are subsequently fused using a preset weight $\gamma$ as the social embeddings $\SM$ of users:
\begin{equation}\label{eq:social-fusion}
\SM = \gamma\cdot{\GM} + (1-\gamma)\cdot{\LM}.
\end{equation}
In the meantime, another encoder is used to generate embeddings $\BM$ of users that capture the collaborative patterns in $\C$ through {\em user-based collaborative encoding} (UCE).
Based thereon, \algo{} additionally introduces the {\em feature mutual exclusion} (FME) between $\SM$ and $\BM$, thereby reducing feature redundancy and learning distinct features for better representations, before combining them using a weighting factor $\beta$ as the final user embeddings in Eq.~\eqref{eq:emb}.
\begin{equation}\label{eq:emb}
\UM = \beta\cdot{\SM} + (1-\beta)\cdot {\BM}.
\end{equation}
Eventually, user embeddings $\UM\in \mathbb{R}^{|\U|\times d}$ and community embeddings $\CM\in \mathbb{R}^{|\V|\times d}$ will be trained via optimizing our joint loss function $\mathcal{L}_{\textnormal{BPR}}+\theta\cdot\mathcal{L}_{\textnormal{KL}}$.

\begin{algorithm}[!t]
\caption{Constructing $\GM$}\label{alg:GSP}
\KwIn{Social network $\G$, initial user embeddings $\UM^{\circ}$}
\Parameter{The number $T$ of iterations, and coefficient $\alpha$}
\KwOut{Embeddings $\GM$}
$\GM^\prime \gets \UM^{\circ},\ \GM\gets \GM^\prime$\;
\For{$t\gets 1$ \KwTo $T$}{
$\GM^{\prime} \gets \frac{\alpha}{(1-\alpha)}\cdot \left(\NAM \GM^{\prime} - \frac{\sqrt{\dvec}}{|\EDG|}\cdot \sqrt{\dvec}^{\top}\GM^{\prime} \right)$\;
$\GM \gets \GM + \GM^{\prime}$\;
}
\end{algorithm}

\subsection{Social Structural Pattern Encoding}
In this section, we elaborate on our SMM and SCA modules for extracting and encoding the community-aware global and local structural features underlying the social network $\G$.

\stitle{Social Modularity Maximization}
Firstly, recall that the global social structural patterns refer to the high internal connectivity and low external connectivity among intra-community users. To comply with this, an ideal complete user-community membership matrix $\DeltaM\in \{0,1\}^{|\U|\times |\V|}$ should maximize the normalized modularity $Q(\NAM)$ defined in Eq.~\eqref{eq:norm_modularity}, which measures the overall internal and external connectivity of the communities over $\G$.
As such, we can formulate the following optimization objective:
\begin{equation}\label{eq:global-obj}
\max_{\DeltaM}\alpha\cdot Q(\NAM) - (1-\alpha)\cdot \|\DeltaM - \YM\|_F^2,
\end{equation}
where $\alpha$ is a coefficient balancing the modularity term and fitting term $\|\DeltaM - \YM\|_F^2$ that measures the distance between target membership matrix $\DeltaM$ and observed memberships in $\YM$.

\begin{lemma}\label{lem:modularity}
$Q(\NAM) = \texttt{trace}\left(\DeltaM^{\top}\left(\NAM- \frac{\sqrt{\dvec}\sqrt{\dvec}^{\top}}{|\EDG|}\right)\DeltaM\right)$ where $\sqrt{\dvec}$ is the element-wise square root of degree vector $\dvec$.
\end{lemma}

Let $\GM$ be the embedding vectors that can be transformed into the target user-community memberships in $\DeltaM$. Building on the optimization objective in Eq.~\eqref{eq:global-obj} and the matrix trace form of $Q(\NAM)$ stated in our Lemma~\ref{lem:modularity}\footnote{Missing proofs can be found in Appendix~\ref{sec:proof} of our technical report~\cite{techreport}.}, the optimization objective towards learning $\GM$ in SMM can be formulated as in Eq.~\eqref{eq:SMM-obj}.
\begin{small}
\begin{equation}\label{eq:SMM-obj}
\max_{\GM}\alpha\cdot \texttt{trace}\left(\GM^{\top}\left(\NAM-\frac{\sqrt{\dvec}\sqrt{\dvec}^{\top}}{|\EDG|}\right)\GM\right) - (1-\alpha)\cdot \|\GM - \UM^{\circ}\|_F^2
\end{equation}
\end{small}
The goal is to maximize the modularity of community memberships derived from embeddings $\GM$ (the first term), while enforcing $\GM$ close to the initial user embeddings via the second term.

\begin{theorem}\label{lem:SMM-sol}
The closed-form solution to the optimization problem in Eq.~\eqref{eq:SMM-obj} is $\GM = \left( \IM - \frac{\alpha}{1-\alpha} \left(\NAM - \frac{\sqrt{\dvec}\sqrt{\dvec}^{\top}}{|\EDG|}\right) \right)^{-1}\UM^{\circ}$.
Particularly, when $\alpha < \frac{1}{3}$, $\GM = \sum_{t=0}^{\infty}{\frac{\alpha^t}{(1-\alpha)^t}\left(\NAM- \frac{\sqrt{\dvec}\sqrt{\dvec}^{\top}}{|\EDG|}\right)^t}\UM^{\circ}$.
\end{theorem}

Our theoretical analysis in Theorem~\ref{lem:SMM-sol} pinpoints that an optimal $\GM$ to the objective in Eq.~\eqref{eq:SMM-obj} can be obtained using the iterative feature propagation in Theorem~\ref{lem:SMM-sol} when $\alpha<\frac{1}{3}$. 
Since it is infeasible to sum up the infinite series, we can estimate $\GM$ with a maximum number $T$ of iterations by
\begin{small}
\begin{equation}\label{eq:SMM}
\GM = \sum_{t=0}^{T}{\frac{\alpha^t}{(1-\alpha)^t}\left(\NAM- \frac{\sqrt{\dvec}\sqrt{\dvec}^{\top}}{|\EDG|}\right)^t}\UM^{\circ}.
\end{equation}
\end{small}

Along this line, Algorithm~\ref{alg:GSP} displays the pseudo-code of calculating $\GM$ in an incremental and decoupled fashion, which can achieve a linear time complexity.

\begin{table}[!t]
\centering
\renewcommand{\arraystretch}{1.2}
\begin{small}
\caption{Neighborhood-based Social Closeness Measures.}\vspace{-3mm} \label{tab:metrics}
\resizebox{\columnwidth}{!}{%
\begin{tabular}{l|c|c}
\hline
{\bf Measure} &  {\bf $p(u_i,u_j)$} & {\bf $f(\AM)$}\\ \hline
  Common Neighbors (CN)   & $|\N_{\G}(u_i)\cap \N_{\G}(u_j)|$ & $\AM$ \\
  Adamic-Adar Index (AAI)~\cite{adamic2003friends} & $\underset{u\in \N_{\G}(u_i)\cap \N_{\G}(u_j)}{\sum}{\frac{1}{\log{d(u)}}}$ & $\AM\DM_{\log{}}^{-\frac{1}{2}}$\\
  Resource Allocation Index (RAI)~\cite{zhou2009predicting} & $\underset{u\in \N_{\G}(u_i)\cap \N_{\G}(u_j)}{\sum}{\frac{1}{d(u)}}$ & $\AM\DM^{-\frac{1}{2}}$ \\
  Salton Index (SI)~\cite{salton1983introduction} & $\frac{|\N_{\G}(u_i)\cap \N_{\G}(u_j)|}{\sqrt{\dvec_i}\sqrt{\dvec_j}}$ & $\DM^{-\frac{1}{2}}\AM$ \\
Leicht-Holme-Newman Index (LHNI)~\cite{leicht2006vertex} & $\frac{|\N_{\G}(u_i)\cap \N_{\G}(u_j)|}{{\dvec_i\cdot \dvec_j}}$ & $\PM$ \\
	 \hline

\end{tabular}%
}
\end{small}
\vspace{-1mm}
\end{table}

\stitle{Social Closeness Aggregation} SCA aims to capture the local social patterns, which implies that users with more common friends, i.e., stronger {\em neighborhood-based social closeness} (NSC), should coexist in more communities. Mathematically, the likelihood that a social user $u_i$ joins the community $c_k$ can be evaluated by
\begin{small}
\begin{equation*}
\DeltaM_{i,k} = \sum_{u_j\in \U}{\frac{p(u_i,u_j)}{\dvec_i}\cdot \YM_{j,k}},
\end{equation*}
\end{small}
where $p(u_i,u_j)$ stands for the NSC between social users $u_i$ and $u_j$, and $\frac{1}{\dvec_i}$ is to average the closeness values from the perspective of user $u_i$.
Building upon this idea, the embedding $\LM_i$ of user $u_i$ can be obtained via an aggregation of other users' initial embeddings:
\begin{small}
\begin{equation}
\LM_{i} = \sum_{u_j\in \U}{\frac{p(u_i,u_j)}{\dvec_i}\cdot \UM^{\circ}_j}.
\end{equation}
\end{small}
Intuitively, this formulation enforces the embeddings of users with strong NSC to be close, and thus, are more likely to be grouped into the same communities.

\begin{theorem}\label{lem:metrics}
For any closeness $p(u_i,u_j)$ in Table~\ref{tab:metrics}, $p(u_i,u_j) = f(\AM)_i\cdot f(\AM)_j^\top$, where $f(\AM)$ is its corresponding transformation of $\AM$.
\end{theorem}

In Table~\ref{tab:metrics}, we list five classic NSC measures, CN, AAI, RAI, SI, and LHNI, for social networks. Theorem~\ref{lem:metrics} unifies all of them into a dot product form, which enables us to compute the embedding vectors using a neat matrix multiplication form as follows: 
\begin{equation}\label{eq:SCF}
\LM = \DM^{-1} f(\AM) \cdot f(\AM)^\top \UM^{\circ}.
\end{equation}

In SCA, we choose the closeness metric RAI~\cite{zhou2009predicting} for computing $\LM$, which is more empirically effective and robust as revealed by our experiments on real datasets in Section~\ref{sec:exp-ablation}.

\subsection{User-based Collaborative Encoding}
Aside from social structural patterns from $\G$, UCE is to capitalize on the rich collaborative signals embodied in CMN $\C$. Recall that in the conventional {\em collaborative filtering}~\cite{su2009survey}, for any user $u_i\in \U$ and community $c_k\in \V$, the rating of $u_i$ on $c_k$, i.e., $u_i$'s probability of joining $c_k$, can be calculated by 
\begin{small}
\begin{equation*}
\DeltaM_{i,k} = \sum_{u_j\in \U}{s(u_i,u_j)\cdot \YM_{j,k}},
\end{equation*}
\end{small}
where $s(u_i,u_j)$ represents the {\em preference similarity} between users $u_i$ and $u_j$ and is computed based on their known community memberships in $\YM$. Following this paradigm, we construct the collaborative embedding $\BM_{i}$ of user $u_i$ via
\begin{small}
\begin{equation}\label{eq:UCF}
\BM_{i} = \sum_{u_j\in \U}{s(u_i,u_j)\cdot \UM^{\circ}_{j}}.
\end{equation}
\end{small}

More precisely, let $\hat{\YM}$ be the normalized and standardized version of $\YM$ whose $(u_i,c_k)$ entry is
\begin{small}
\begin{equation}\label{eq:Yhat}    
\hat{\YM}_{i,k} = \frac{\YM_{i,k}}{\sqrt{\dtvec_i}\cdot\sqrt{\svec_k}} - \sqrt{\frac{\dtvec_i}{|\M|}} \cdot \sqrt{\frac{\svec_k}{|\M|}}.
\end{equation}
\end{small}
The first term averages the membership $\YM_{i,k}$ with the consideration of $u_i$'s community engagement and $c_k$'s popularity, whilst $\sqrt{\frac{\dtvec_i}{|\M|}} \cdot \sqrt{\frac{\svec_k}{|\M|}}$ acts as the global preference biases of $u_i$ and $c_k$.
Notably, $\hat{\YM}_{i,k}$ is essentially an element of the {\em bipartite modularity}~\cite{barber2007modularity} of $\C$.
Based thereon, the preference similarity $s(u_i,u_j)$ between users $u_i$ and $u_j$ is defined as
\begin{equation}\label{eq:preference-sim}
s(u_i,u_j) = \hat{\YM}_i \hat{\YM}_j^{\top}.
\end{equation}
Accordingly, the collaborative embeddings of users are then calculated using $\BM = \hat{\YM} \hat{\YM}^{\top} \UM^\circ$.

\begin{lemma}\label{lem:B-bound}
$\textstyle \forall{u_i, u_j}\in \U$, 
$-\frac{ \sqrt{\dtvec_i}\sqrt{\dtvec_j}}{|\M|}\le s(u_i,u_j) \le 1- \frac{ \sqrt{\dtvec_i}\sqrt{\dtvec_j}}{|\M|}$.
\end{lemma}
Notably, the formulation in Eq.~\eqref{eq:preference-sim} renders the preference similarity $s(u_i,u_j)\ \forall{u_i,u_j\in \U}$ fall into a better range given in Lemma~\ref{lem:B-bound}, which accounts for the preference biases for different user pairs, and can be more accurate in discriminating similar and dissimilar users with a bound of $-1$ and $1$.

\subsection{Feature Mutual Exclusion}\label{sec:FME}
As remarked in Section~\ref{sec:overview}, one design recipe of \algo{} is to leverage the complementary nature of social structures and collaborative structures. 
However, due to the high consistency between social connections and user preferences in communities \cite{mcpherson2001birds_homophily}, a large overlap might exist in social and collaborative features $\SM$ and $\BM$, which in turn engenders severe feature redundancy and dilutes the complementary capability.

To remedy this issue, we resort to minimizing the {\em Hilbert-Schmidt Independence Criterion} (HSIC)\footnote{The detailed definition is deferred to Appendix~\ref{sec:HSIC} of our technical report~\cite{techreport}.} \cite{gretton2005measuring_HSIC} of $\SM$ and $\BM$ to achieve the mutual exclusion of their features for more expressive user embeddings. Particularly, $\texttt{HSIC}(\SM, \BM)$ is defined based on the cross-covariance and essentially measures the statistical dependence between $\SM$ and $\BM$~\cite{gretton2005measuring}.
Due to practicality, we calculate the empirical HSIC~\cite{gretton2005measuring_HSIC} between $\SM$ and $\BM$, which can be further simplified as follows:
\begin{equation}\label{eq:HSIC-fast}
\texttt{HSIC}(\SM, \BM) = \texttt{trace}(\SM^{\top}\BM\BM^{\top}\SM) = \|\SM^{\top}\BM\|_F^2,
\end{equation}
if we choose the linear kernel as the Gram matrix in it and each row in $\SM$ and $\BM$ is $L_2$ normalized.

For ease of exposition, we denote by $\SM^{\circ}$ the initial social embeddings output by SMM and SCA, and by $\BM^\circ$ the initial collaborative embeddings returned by UCE. The optimization goal of FME thus can be formulated as the minimization of $\mathcal{O}_{\textnormal{ME}}$:
\begin{small}
\begin{equation}\label{eq:ZR-FME-obj}
\mathcal{O}_{\textnormal{ME}} = \|{\SM} - \SM^\circ\|_F^2 + \|{\BM}-\BM^\circ\|_F^2 + \lambda \cdot \texttt{HSIC}(\SM, \BM).
\end{equation}
\end{small}
The hyper-parameter $\lambda$ is used to adjust the impact of $\texttt{HSIC}(\SM, \BM)$, i.e., the feature mutual exclusion.
By setting its derivative w.r.t. $\SM$ and $\BM$ to zero, we have
\begin{small}
\begin{align*}
\frac{\partial \mathcal{O}_{\textnormal{ME}}}{\partial {\SM}} &= {\SM}-\SM^\circ + \lambda \cdot {\BM} {\BM}^\top {\SM}=0,\quad \frac{\partial \mathcal{O}_{\textnormal{ME}}}{\partial {\BM}} &= {\BM}-\BM^\circ + \lambda \cdot {\SM} {\SM}^\top {\BM}=0,
\end{align*}
\end{small}
which leads to the following rules for iteratively updating $\SM$ and $\BM$:
\begin{small}
\begin{equation}\label{eq:FME}
\begin{split}
{\SM}^{(i+1)} = \SM^\circ - \lambda\cdot {\BM}^{(i)}{\BM}^{(i)\top} {\SM}^{(i)},\quad {\BM}^{(i+1)} = \BM^\circ - \lambda\cdot {\SM}^{(i)}{\SM}^{(i)\top}{\BM}^{(i)},
\end{split}
\end{equation}
\end{small}
where ${\SM}^{(0)}={\SM}^\circ$ and ${\BM}^{(0)}={\BM}^\circ$, and every row in ${\SM}^\circ$ and ${\BM}^\circ$ is normalized to a unit vector. It is worth mentioning that this normalization is not only to conform to the requirements for $\SM$ and $\BM$ in Eq.~\eqref{eq:HSIC-fast}, but also can circumvent the exploding gradients caused by the significant magnitudes of ${\SM}^{(i)}{\SM}^{(i)\top}$ and ${\BM}^{(i)}{\BM}^{(i)\top}$ in Eq.~\eqref{eq:FME}.

\begin{algorithm}[!t]
\caption{The training of \algo{}}\label{alg:main}
\KwIn{$\G=(\U,\EDG)$ and $\C=(\U\cup \V,\M)$.}
\Parameter{$\alpha, \beta, \gamma, \lambda$}
\KwOut{User and community embeddings $\UM$ and $\CM$.}
Initialize embedding vectors $\UM^{\circ}$ and $\CM$\;
\For{each epoch}{
Invoking Algorithm~\ref{alg:GSP} to compute $\GM$\;
Compute $\LM$ according to Eq.~\eqref{eq:SCF}\;
$\SM \gets \gamma\cdot\GM + (1-\gamma)\cdot\LM$ \;
Compute $\BM$ according to Eq.~\eqref{eq:UCF}\;
Update ${\SM}$, ${\BM}$ according to Eq.~\eqref{eq:FME}\;
$\UM \gets \beta\cdot{\SM} + (1-\beta)\cdot {\BM}$\;
Compute the joint loss $\mathcal{L}$ according to Eq.~\eqref{loss:all}\;
Update $\UM^\circ$ and $\CM$ by the backpropagation\;
}
\end{algorithm}

\subsection{Prediction and Optimization}\label{sec:training}

After deriving user embeddings $\UM$ and community embeddings $\CM$, \algo{} makes community recommendations to users by calculating the prediction score $\hat{y}_{i,k}$ of user $u_i$ to community $c_k$ by the inner product of their corresponding embeddings:
\begin{equation}
\hat{y}_{i,k} = \UM_i\CM_{k}^{\top}.
\end{equation}
The top-$K$ communities ranked according to the prediction scores will be used as the final recommendation list for each user.

\stitle{Recommendation Loss}
We adopt the canonical {\em Bayesian Personalized Ranking} (BPR) loss~\cite{rendle2009bpr} for recommendation model training, which is a pairwise loss encouraging the prediction scores of the observed interactions to be higher than those that are unobserved:
\begin{small}
\begin{equation*}\label{loss:BPR}
\mathcal{L}_{\textnormal{BPR}} = -\sum_{u_i\in \U}{\sum_{c_k\in \M(u)}{\sum_{c_j\notin \M(u)}}{\ln{\sigma(\hat{y}_{i,k}-\hat{y}_{i,j})}}} + \zeta \cdot \left(\|\UM\|^2 + \|\CM\|^2\right),
\end{equation*}
\end{small}
where $\sigma(\cdot)$ stands for the sigmoid activation function and $\zeta$ is the regularization coefficient.

\stitle{Community Detection Loss}
Apart from the recommendation loss, we additionally include a community detection loss for model training. The rationale is that, distinct from the usual items in recommender systems, each community is composed of a cluster of users, and thus, can be characterized by its members in the form of the cluster centroid. Ideally, each cluster centroid $c_k$ is required to be close to its internal users, while being distant from those belonging to other clusters.

Inspired by this, we adopt the classic Kullback-Leibler (KL) divergence loss~\cite{xie2016unsupervised_cluster_loss} for deep clustering, which refines the cluster centroids, i.e., community embeddings, by matching the predicted soft user-community assignments to an auxiliary target distribution.
In mathematical terms, the objective is defined as a KL divergence between the soft assignments $P$ and the target distribution $Q$ as follows:
\begin{small}
\begin{equation}\label{loss:KL}
\mathcal{L}_{\textnormal{KL}} = \texttt{KL}(P\|Q) = \sum_{u_i\in \U}\sum_{c_k\in \V}{p_{i,k}\cdot \log{\frac{p_{i,k}}{q_{i,k}}}},
\end{equation}
\end{small}
where $q_{i,k}\in Q$ represents the soft assignment, i.e., the probability of assigning user $u_i$ to community $c_k$, and $p_{i,k}\in P$ denotes the soft probabilistic target of $q_{i,k}$.
Following~\cite{xie2016unsupervised_cluster_loss}, the Student’s t-distribution is employed as a kernel to measure the similarity between each user and the community, i.e., soft assignment:
\begin{small}
\begin{equation}\label{eq:q}
q_{i,k} = \frac{(1+\|\UM_i - \CM_k\|^2)^{-1}}{\sum_{c_\ell \in \V}(1+\|\UM_i - \CM_\ell\|^2)^{-1}}.
\end{equation}
\end{small}
As for the target distribution $P$, we can simply harness the observed user-community memberships in $\YM$ by applying a normalization, i.e., $p_{i,k} = {\YM_{i,k}}/{\sum_{c_\ell\in \V}{\YM_{i,\ell}}}$.

\stitle{Optimization} Overall, the training objective is a combination of the recommendation and community detection loss functions:
\begin{equation}\label{loss:all}
\mathcal{L}_{\textnormal{BPR}}+\theta \cdot\mathcal{L}_{\textnormal{KL}},
\end{equation}
where $\theta$ is a coefficient for this joint loss.
In Algorithm~\ref{alg:main}, we summarize the main training process of \algo{}.
Notice that the training parameters are solely the initial user embeddings $\UM^\circ$ and community embeddings $\CM$. As such, \algo{} also enjoys efficient model training as matrix factorization and many others~\cite{he2020lightgcn}.

\subsection{Model Complexity Analysis}
The time cost of \algo{} mainly lies in the four components: SMM, SCA, UCE, and FME. 
According to Lines 2-4 in Algorithm~\ref{alg:GSP}, each iteration conducts a sparse matrix multiplication that takes $O(|\EDG|\cdot d)$ time where $d$ stands for the embedding dimension. The total computational cost needed by SMM is thus $O(|\EDG|\cdot dT)$.
By reordering the matrix multiplications in Eq.~\eqref{eq:SCF} and Eq.~\eqref{eq:UCF} and using sparse operations, SCA and UCE can be done in $O(|\EDG|\cdot d)$ and $O(|\M|\cdot d)$ time, respectively. 
In FME, we conduct the updating operations in Eq.~\eqref{eq:FME} for merely one iteration, wherein $\lambda\cdot {\BM}^{(i)}{\BM}^{(i)\top} {\SM}^{(i)}$ and $\lambda\cdot {\SM}^{(i)}{\SM}^{(i)\top}{\BM}^{(i)}$ can be reordered as $\lambda\cdot {\BM}^{(i)}\cdot ({\BM}^{(i)\top} {\SM}^{(i)})$ and $\lambda\cdot {\SM}^{(i)}\cdot ({\SM}^{(i)\top}{\BM}^{(i)})$, respectively, leading to a complexity of $O(|\U|\cdot d^2)$.
In total, the time complexity for SMM, SCA, UCE, and FME can be bounded by $O(|\EDG|\cdot dT + |\M|\cdot d +|\U|\cdot d^2)$. 
As for the training, the complexity is the same as the standard matrix factorization model using the BPR loss.

In addition to input data $\G$ and $\C$, \algo{} materializes $|\U|\times d$ embedding vectors $\GM$, $\LM$, $\SM$, $\BM$, and $\UM$ for users, and community embeddings $\CM \in \mathbb{R}^{|\V|\times d}$.
Therefore, the space complexity of \algo{} is bounded by $O(|\EDG|+|\M|+(|\U|+|\V|)\cdot d)$, which is linear to the input and output.

\section{Experiments}
This section experimentally evaluates \algo{} against 9 baselines in terms of community recommendation on 6 real datasets, and conducts related ablation studies and hyperparameter analyses.
All experiments are conducted on a Linux machine with an NVIDIA Ampere A100 GPU (80 GB memory), AMD EPYC 7513 CPUs (2.6 GHz), and 1TB RAM. The codes and datasets are made publicly available at \url{https://github.com/HKBU-LAGAS/CASO}. 

\begin{table}[!t]
\renewcommand{\arraystretch}{0.9}
\centering
\caption{Dataset statistics.}
\label{tab:stats_brief}
\vspace{-2.5ex}
\begin{small}
\begin{tabular}{c|c|c|c|c|c}
\hline
{\bf Dataset} & {\bf $|\U|$} & {\bf $|\EDG|$} & {\bf $|\V|$} & {\bf $|\M|$} & {\bf $|\M|/|\U|$}\\ \hline
{\em BlogCatalog}~\cite{meng2019co-minidata} & 5,196 & 171,743 & 6 & 5,196 & 1.0000\\
{\em Flickr}~\cite{meng2019co-minidata} & 7,575 & 479,476 & 9 & 7,575 & 1.0000\\
{\em Deezer-HR}~\cite{rozemberczki2019gemsec-Deezer} & 54,573 & 498,202 & 84 & 343,665 & 6.2973\\
{\em Deezer-RO}~\cite{rozemberczki2019gemsec-Deezer} & 41,773 & 125,826 & 84 & 252,132 & 6.0358\\
{\em DBLP}~\cite{yang2012defining-largedata} & 317,080 & 1,049,866 & 5,000 & 112,228 & 0.3539\\
{\em Youtube}~\cite{yang2012defining-largedata} & 1,134,890 & 2,987,624 & 5,000 & 72,959 & 0.0643\\
\hline
\end{tabular}
\end{small}
\label{tab:datasets}
\vspace{-2ex}
\end{table}

\subsection{Experimental Setup}\label{sec:exp-set}
\stitle{Datasets}
Table~\ref{tab:stats_brief} summarizes the six datasets used in the experiments, including {\em BlogCatalog}~\cite{meng2019co-minidata}, {\em Flickr}~\cite{meng2019co-minidata}, {\em Deezer-HR}~\cite{rozemberczki2019gemsec-Deezer}, {\em Deezer-RO}~\cite{rozemberczki2019gemsec-Deezer}, {\em DBLP}~\cite{yang2012defining-largedata}, and {\em Youtube}~\cite{yang2012defining-largedata}, each of which is associated with a social network and a CMN collected from open-source platforms/projects.
For each dataset, we randomly sample 80\% community memberships of each user for training and validation, 
while the rest 20\% is used for testing. 
A negative community is randomly sampled for each positive instance to form the training set.

\stitle{Baselines and Hyperparameters}
We evaluate \algo{} against 9 competitors, including 6 recent social recommendation models~\cite{yang2024GBSR,wu2019diffnet,wu2020diffnet++,yu2021SEPT,yu2021MHCN,yang2024GBSR}, as well as 3 collaborative filtering methods~\cite{koren2008SVD++,rendle2009bpr,he2020lightgcn}, a matrix factorization method~\cite{koren2008SVD++}, and a hypergraph convolutional network model~\cite{yu2021MHCN} for recommendation:
\begin{itemize}[leftmargin=*]
\item {\texttt{SVD++}}~\cite{koren2008SVD++} combines the advantages of matrix factorization and neighborhood-based collaborative filtering methods
\item {\texttt{BPR}}~\cite{rendle2009bpr} is a popular recommendation model and classic latent factor-based technique. 
\item {\texttt{LightGCN}}~\cite{he2020lightgcn} is an efficient graph-based collaborative filtering model, which simplifies GCNs by removing the redundant components for recommendation.
\item {\texttt{LightGCN-S}}~\cite{yang2024GBSR} extends LightGCN to social recommendation, where the information of each user is aggregated from their interacted items and linked social users.
\item {\texttt{DiffNet}}~\cite{wu2019diffnet} is a classic graph-based social recommendation method that models the recursive dynamic social diffusions.
\item {\texttt{DiffNet++}}~\cite{wu2020diffnet++} is an improved model of DiffNet that incorporates a multi-level attention network to astutely capture users' preferences for different graph sources. 
\item {\texttt{SEPT}}~\cite{yu2021SEPT} is a general socially-aware self-supervised multi-view co-training framework for recommendation. 
\item {\texttt{MHCN}}~\cite{yu2021MHCN} is a hypergraph convolutional network-based social recommendation method by leveraging diverse types of high-order user relations.
\item {\texttt{GBSR}}~\cite{yang2024GBSR} is a model-agnostic social denoising framework using the information bottleneck principle to steer the denoising.
\end{itemize}
We run the codes collected from their respective authors with parameter settings via grid search.
Following common practice, the embedding size is set to $64$ in all evaluated models.
In \algo{}, we fix $\alpha=0.33$, $\gamma=0.3$, and $T=2$, while $\lambda$, $\beta$, and $\theta$ are tuned in the ranges of $[0.001, 1]$, $[0.1, 1]$, and $[0.001, 1]$, respectively. We employ the Adam~\cite{kingma2014adam} optimizer with a learning rate of 0.01 and a mini-batch size of 2048. A maximum number of $1000$ epochs, together with early stopping and validation strategies, is used.

Due to space constraints, we refer interested readers to Appendix~\ref{sec:exp-add}~\cite{techreport} for the details of datasets and hyperparameters.

\begin{table*}[!t]
\centering
\renewcommand{\arraystretch}{0.85}
\caption{Overall recommendation performance. Best is highlighted in bold and runner-up \underline{underlined}.}
\label{tab:performance}
\vspace{-2.5ex}
\addtolength{\tabcolsep}{-0.2em}
\resizebox{\textwidth}{!}{%
    \begin{tabular}{c c c c c c c c c c c c c}
        \toprule
         \multirow{2}{*}{\bf Method} & \multicolumn{4}{c}{\em \bf BlogCatalog} & \multicolumn{4}{c}{\em \bf Flickr} & \multicolumn{4}{c}{\bf Deezer-HR} \\
        \cmidrule(lr){2-5} \cmidrule(lr){6-9} \cmidrule(lr){10-13} 
        & Recall@3 & NDCG@3 & Recall@5 & NDCG@5 & Recall@3 & NDCG@3 & Recall@5 & NDCG@5 & Recall@3 & NDCG@3 & Recall@5 & NDCG@5 \\
        \midrule
        \texttt{SVD++}~\cite{koren2008SVD++} & 0.5052 & 0.3587 & 0.8331 & 0.4854  & 0.3373 & 0.2447 & 0.5599 & 0.3264 & 0.1864 & 0.1694 & 0.3187 & 0.2147  \\
        \texttt{BPR}~\cite{rendle2009bpr} & 0.5069 & 0.3570 & 0.8387 & 0.4921  & 0.3358 & 0.2362 & 0.5612 & 0.3299 & 0.5213 & 0.5169 & 0.6467 & 0.5605  \\
        \texttt{LightGCN}~\cite{he2020lightgcn} & 0.5035 & 0.3590 & 0.8322 & 0.4920 & 0.3352 & 0.2364 & 0.5566 & 0.3277 & 0.5030 & 0.5018 & 0.6308 & 0.5462  \\
        \texttt{LightGCN-S}~\cite{yang2024GBSR} & 0.9026 & 0.7713 & 0.9831 & 0.8037 & 0.6705 & 0.5447 & 0.8352 & 0.6124 & 0.5262 & 0.5194 & 0.6508 & 0.5638  \\
        \texttt{DiffNet}~\cite{wu2019diffnet} & 0.8962 & 0.7875 & 0.9842 & 0.8232 & 0.7532 & 0.6309 & 0.8855 & 0.6838 & 0.3617 & 0.3351 & 0.4813 & 0.3832 \\
        \texttt{DiffNet++}~\cite{wu2020diffnet++} & 0.8856 & 0.7753 & 0.9747 & 0.8131 & 0.6657 & 0.5310 & 0.8257 & 0.5974 & 0.3578 & 0.3240 & 0.4778 & 0.3713 \\
        \texttt{SEPT}~\cite{yu2021SEPT} & 0.5106 & 0.3664 & 0.8251 & 0.4991 & 0.3950 & 0.3086 & 0.5741 & 0.3774 & 0.5331 & 0.5206 & 0.6576 & 0.5635  \\
        \texttt{MHCN}~\cite{yu2021MHCN} & \underline{0.9419} & \underline{0.8379} & \underline{0.9915} & \underline{0.8609} & \underline{0.8069} & \underline{0.6789} & \textbf{0.9381} & \underline{0.7277} & 0.1346 & 0.1144 & 0.2229 & 0.1532  \\
        \texttt{GBSR}~\cite{yang2024GBSR} & 0.8530 & 0.7426 & 0.9646 & 0.7873 & 0.6004 & 0.4826 & 0.7752 & 0.5537 & \underline{0.5342} & \underline{0.5234} & \underline{0.6718} & \underline{0.5738} \\
        \midrule
        \algo{} & \textbf{0.9517} & \textbf{0.8535} & \textbf{0.9962} & \textbf{0.8741} & \textbf{0.8191} & \textbf{0.6976} & \underline{0.9333} & \textbf{0.7451} & \textbf{0.5602} & \textbf{0.5446} & \textbf{0.6983} & \textbf{0.5949} \\
        Improv. & \textbf{1.046\%} & \textbf{1.868\%} & \textbf{0.466\%} & \textbf{1.527\%} & \textbf{1.522\%} & \textbf{2.755\%} & -0.507\% & \textbf{2.389\%} & \textbf{4.856\%} & \textbf{4.054\%} & \textbf{3.939\%} & \textbf{3.676\%} \\
        \bottomrule
    \end{tabular}
}
\resizebox{\textwidth}{!}{%
    \begin{tabular}{c c c c c c c c c c c c c}
        \toprule
         \multirow{2}{*}{\bf Method} & \multicolumn{4}{c}{\em \bf Deezer-RO} & \multicolumn{4}{c}{\em \bf DBLP} & \multicolumn{4}{c}{\bf Youtube} \\
        \cmidrule(lr){2-5} \cmidrule(lr){6-9} \cmidrule(lr){10-13} 
        & Recall@3 & NDCG@3 & Recall@5 & NDCG@5 & Recall@3 & NDCG@3 & Recall@5 & NDCG@5 & Recall@3 & NDCG@3 & Recall@5 & NDCG@5 \\
        \midrule
        \texttt{SVD++}~\cite{koren2008SVD++} & 0.1127 & 0.0743 & 0.2174 & 0.1189 & 0.0006 & 0.0004 & 0.0087 & 0.0036 & 0.0003 & 0.0003 & 0.0006 & 0.0004 \\
        \texttt{BPR}~\cite{rendle2009bpr} & 0.5297 & 0.5164 & 0.6677 & 0.5665 & 0.1666 & 0.1256 & 0.2339 & 0.1532 & 0.1059 & 0.0903 & 0.1414 & 0.1048 \\
        \texttt{LightGCN}~\cite{he2020lightgcn} & 0.5152 & 0.5105 & 0.6371 & 0.5518 & 0.1572 & 0.1332 & 0.1887 & 0.1447 & 0.1908 & 0.1680 & 0.2346 & 0.1861 \\
        \texttt{LightGCN-S}~\cite{yang2024GBSR} & 0.5337 & 0.5162 & 0.6609 & 0.5624 & 0.8993 & 0.8215 & 0.9366 & 0.8372 & 0.5122 & 0.4501 & 0.5929 & 0.4838 \\
        \texttt{DiffNet}~\cite{wu2019diffnet} & 0.3569 & 0.3263 & 0.4964 & 0.3840 & 0.7326 & 0.6556 & 0.7993 & 0.6836 & 0.3097 & 0.2618 & 0.3775 & 0.2907 \\
        \texttt{DiffNet++}~\cite{wu2020diffnet++} & 0.3511 & 0.3187 & 0.4985 & 0.3800& 0.8130 & 0.7633 & 0.8436 & 0.7763 & 0.3852 & 0.3435 & 0.4445 & 0.3692 \\
        \texttt{SEPT}~\cite{yu2021SEPT} & 0.5357 &  \underline{0.5243} & 0.6644 &  \underline{0.5688} & 0.3925 & 0.3386 & 0.4231 & 0.3514 & 0.2133 & 0.1858 & 0.2603 & 0.2055 \\
        \texttt{MHCN}~\cite{yu2021MHCN} &  \underline{0.5351} & 0.5165 & 0.6628 & 0.5612 & 0.8663 & 0.8024 & 0.9037 & 0.8170 & 0.3709 & 0.3237 & 0.4538 & 0.3543 \\
        \texttt{GBSR}~\cite{yang2024GBSR} & 0.5323 & 0.5160 &  \underline{0.6750} & 0.5683 & \underline{0.9039} & \underline{0.8251} & \textbf{0.9424} & \underline{0.8398} & \underline{0.5540} & \underline{0.4870} & \underline{0.6348} & \underline{0.5203} \\
        \midrule
        \algo{} & \textbf{0.5643} & \textbf{0.5423} & \textbf{0.7019} & \textbf{0.5906} & \textbf{0.9146} & \textbf{0.8656} & \underline{0.9415} & \textbf{0.8768} & \textbf{0.5681} & \textbf{0.5132} & \textbf{0.6413} & \textbf{0.5435} \\
        Improv. & \textbf{5.335\%} & \textbf{3.434\%} & \textbf{3.984\%} & \textbf{3.837\%} & \textbf{1.182\%} & \textbf{4.904\%} & -0.099\% & \textbf{4.404\%} & \textbf{2.540\%} & \textbf{5.380\%} & \textbf{1.038\%} & \textbf{4.443\%} \\
        \bottomrule
    \end{tabular}
}
\end{table*}

\begin{figure}[!t]
    \centering
    
    \begin{tikzpicture}
        \begin{customlegend}[
            legend entries={w/o SMM, w/o SCA, w/o FME, w/o UCE, w/o $\mathcal{L}_{\textnormal{KL}}$, \algo{}},
            legend columns=3,
            area legend,
            legend style={at={(0.45,1.15)},anchor=north,draw=none,font=\small,column sep=0.4cm}
        ]
            \addlegendimage{preaction={fill, NSCcol1}} 
            \addlegendimage{preaction={fill, NSCcol2}}
            \addlegendimage{preaction={fill, NSCcol3}}
            \addlegendimage{preaction={fill, NSCcol4}}
            \addlegendimage{preaction={fill, NSCcol6}}
            \addlegendimage{preaction={fill, NSCcol5}}
            
        \end{customlegend}
    \end{tikzpicture}
    \vspace{-3mm}
    \\[-\lineskip]
    \subfloat[{\em BlogCatalog}]{%
        \includegraphics[width=0.49\columnwidth]{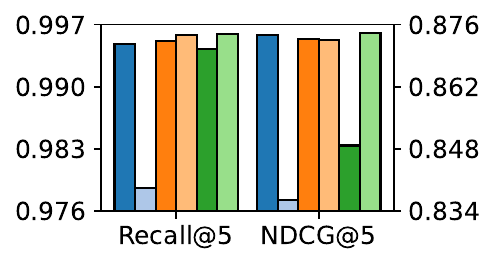}
    }
    \subfloat[{\em Flickr}]{%
        \includegraphics[width=0.49\columnwidth]{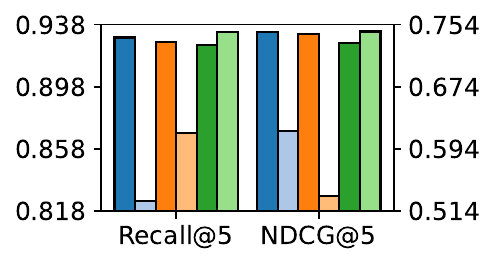}
    }
    \vspace{-3mm}
    \subfloat[{\em Deezer-HR}]{%
        \includegraphics[width=0.49\columnwidth]{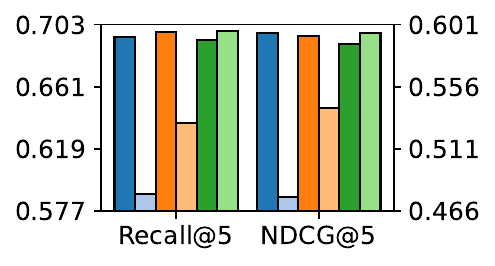} 
    }
    \subfloat[{\em Deezer-RO}]{%
        \includegraphics[width=0.49\columnwidth]{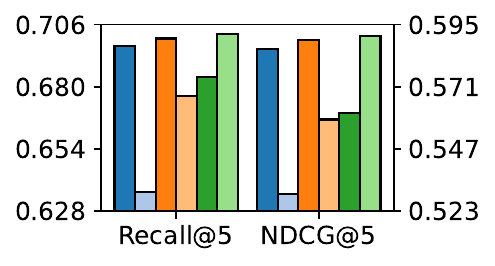} 
    }
    \vspace{-3mm}
    \subfloat[{\em DBLP}]{%
        \includegraphics[width=0.49\columnwidth]{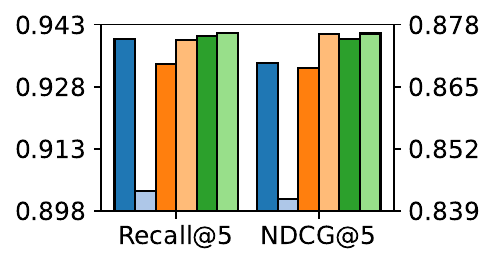}
    }
    \subfloat[{\em Youtube}]{%
        \includegraphics[width=0.49\columnwidth]{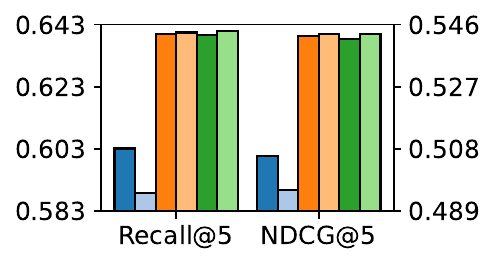}
    }
    \vspace{-1ex}
    \caption{Ablation study.}
    \label{fig:ablation}
\end{figure}

\stitle{Evaluation Protocol}
We adopt two widely used ranking metrics Recall@$K$ and NDCG@$K$, for the evaluation of the top-$K$ ($K=3, 5$) community recommendation performance. Similar to previous works~\cite{krichene2020sampled}, all non-interacted communities are considered as candidates for recommendation to users to avoid biased evaluation.
For each model, we follow 5-fold cross-validation on each dataset and report the results averaged over 10 trials.

\begin{figure}[!t]
    \centering
    
    \begin{tikzpicture}
        \begin{customlegend}[
            legend entries={CN, AAI, RAI, SI, LHNI},
            legend columns=5,
            area legend,
            legend style={at={(0.45,1.15)},anchor=north,draw=none,font=\footnotesize,column sep=0.25cm}
        ]
            \addlegendimage{preaction={fill, NSCcol1}} 
            \addlegendimage{preaction={fill, NSCcol2}}
            \addlegendimage{preaction={fill, NSCcol3}}
            \addlegendimage{preaction={fill, NSCcol4}}
            \addlegendimage{preaction={fill, NSCcol5}}
            
        \end{customlegend}
    \end{tikzpicture}
    \vspace{-3mm}
    
    \subfloat[{\em BlogCatalog}]{%
        \includegraphics[width=0.50\columnwidth]{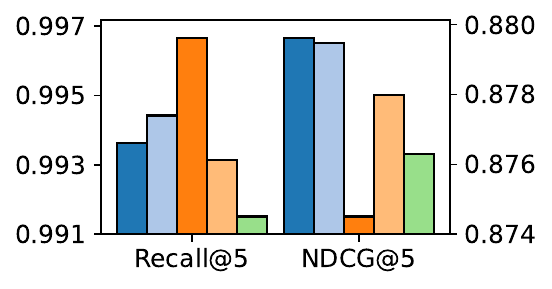}
    }
    \subfloat[{\em Flickr}]{%
        \includegraphics[width=0.49\columnwidth]{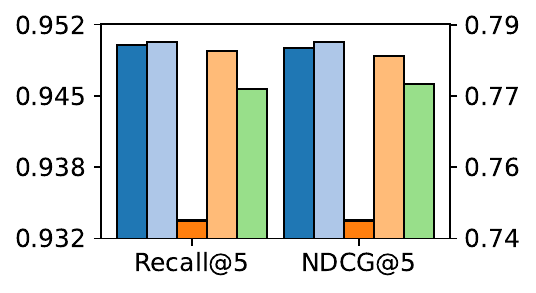}
    }
    \vspace{-3mm}
    \subfloat[{\em Deezer-HR}]{%
        \includegraphics[width=0.50\columnwidth]{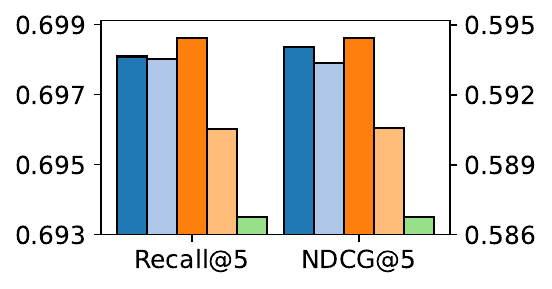} 
    }
    \subfloat[{\em Deezer-RO}]{%
        \includegraphics[width=0.50\columnwidth]{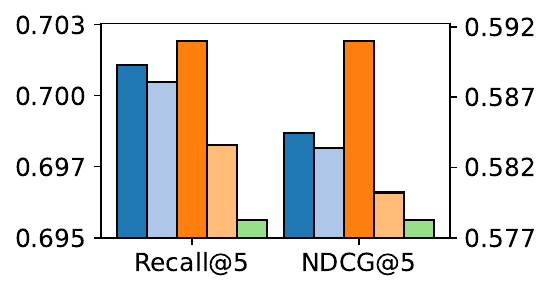} 
    }
    \vspace{-3mm}
    \subfloat[{\em DBLP}]{%
        \includegraphics[width=0.51\columnwidth]{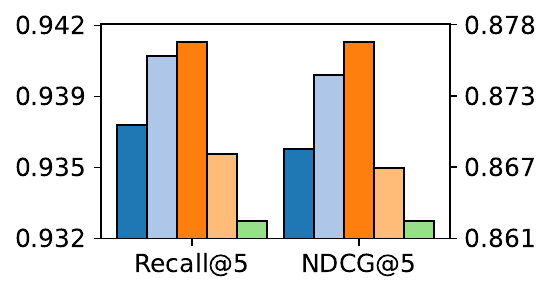}
    }
    \subfloat[{\em Youtube}]{%
        \includegraphics[width=0.48\columnwidth]{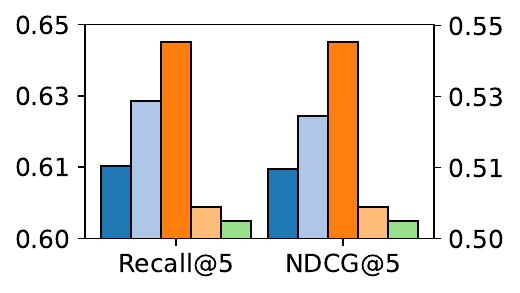}
    }
    \vspace{-2ex}
    \caption{\algo{} with various NSC Measures.}
    \label{fig:NSC}
\end{figure}

\subsection{Recommendation Performance Evaluation}\label{sec:exp-recommend}
Table~\ref{tab:performance} reports the Recall@$K$ and NDCG@$K$ ($K=3, 5$) results of \algo{} and competing models on the six datasets. 
The results under other $K$ values are quantitatively similar, and thus, are deferred to Appendix~\ref{sec:exp-add}~\cite{techreport} for the interest of space.
We can make the following observations from Table~\ref{tab:performance}.

Firstly, our proposed \algo{} almost consistently outperforms all baselines across all tested datasets with conspicuous improvements.
For instance, \algo{} improves the best baselines by considerable gains of 1.5\%, 2.4\%, 3.7\%, and 3.8\% in terms of NDCG@5 on {\em BlogCatalog}, {\em Flickr}, {\em Deezer-HR}, and {\em Deezer-RO}, respectively. 
Particularly, on {\em DBLP} and {\em Youtube} containing millions of social friendships, \algo{} is able to achieve a large margin of $4.9\%$ and $5.4\%$ in NDCG@3, and $4.4\%$ in NDCG@5, respectively.
The improvements attained by \algo{} in recall are also notable, often ranging from $1\%$ to up to $5.3\%$, on all datasets except {\em Flickr} and {\em DBLP}, where \algo{} is comparable to the state-of-the-art models \texttt{MHCN} and \texttt{GBSR} in Recall@5 while being remarkably superior in Recall@3, NDCG@3, and NDCG@5.
The results demonstrate the effectiveness of our \algo{} in the exploitation and integration of social and collaborative information for community recommendation. 

Moreover, compared with simple recommendation methods, i.e., \texttt{SVD++}, \texttt{BPR}, \texttt{LightGCN}, that only rely on the CMN $\C$, \texttt{LightGCN-S} and social recommendation models, e.g., \texttt{DiffNet} and \texttt{MHCN}, generally exhibit higher performance on most datasets, by virtue of the inclusion of social structures in $\G$.
However, these social recommendation models are mainly designed for recommending regular items. Many of them are suboptimal or even fail on datasets like {\em Deezer-HR}, due to the neglect of global social patterns and community structures that are crucial for community recommendation.

\definecolor{pythonBlue}{RGB}{31,119,180}  %
\definecolor{pythonOrange}{RGB}{255,127,14} %
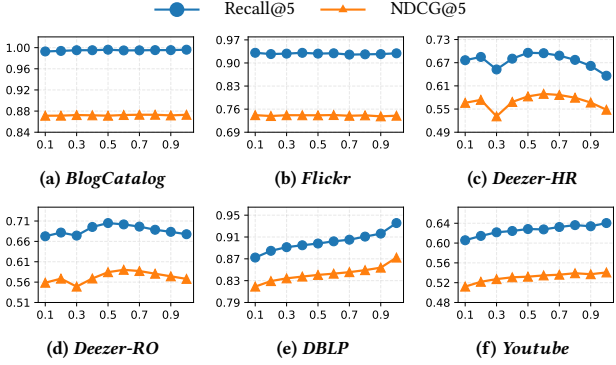
\begin{figure}[!t]
    \centering
    \begin{tikzpicture}
        \begin{customlegend}[
            legend entries={Recall@5, NDCG@5},
            legend columns=3,
            area legend,
            legend style={at={(0.5,1.05)},anchor=north,draw=none,font=\footnotesize,column sep=0.25cm}
        ]
            \addlegendimage{line legend, solid, thick, color=pythonBlue, mark=*, 
            mark options={solid, fill=pythonBlue, scale=1.2}}
            \addlegendimage{line legend, solid, thick, color=pythonOrange, mark=triangle*,
            mark options={solid, fill=pythonOrange, scale=0.6, rotate=0}}
        \end{customlegend}
        \coordinate (legend) at (current bounding box.north);
    \end{tikzpicture}

    \vspace{-4mm}
    \subfloat[{\em BlogCatalog}]{%
        \includegraphics[width=0.32\columnwidth]{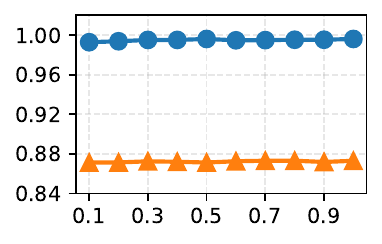}
    }
    \subfloat[{\em Flickr}]{%
        \includegraphics[width=0.32\columnwidth]{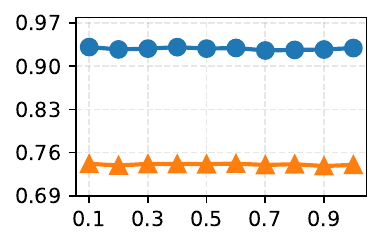}
    }
    \subfloat[{\em Deezer-HR}]{%
        \includegraphics[width=0.32\columnwidth]{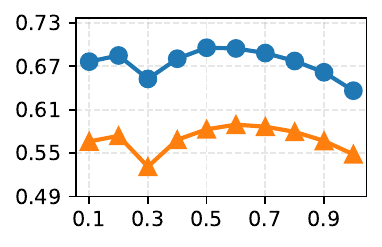} 
    }
    \vspace{-2ex}
    \subfloat[{\em Deezer-RO}]{%
        \includegraphics[width=0.32\columnwidth]{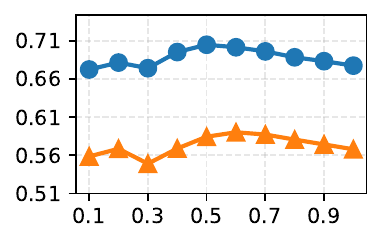} 
    }
    \subfloat[{\em DBLP}]{%
        \includegraphics[width=0.32\columnwidth]{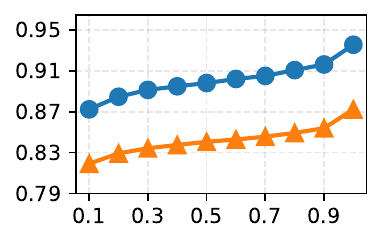}
    }
    \subfloat[{\em Youtube}]{%
        \includegraphics[width=0.32\columnwidth]{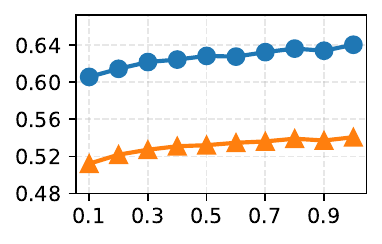}
    }
    \vspace{-1ex}
    \caption{Varying $\beta$ in datasets.}
    \label{fig:vary-beta}

\end{figure}

\subsection{Ablation Study}\label{sec:exp-ablation}
This set of experiments is to investigate the efficacy of the four key components in \algo{}, including SMM, SCA, UCE, and FME, as well as the community detection loss $\mathcal{L}_{\textnormal{KL}}$ and the choice of the NSC measures in SCA.
Fig.~\ref{fig:ablation} depicts the Recall@5 and NDCG@5 results obtained by \algo{} and its five ablated versions on all datasets.

\stitle{w/o SMM}
From Fig. \ref{fig:ablation}, compared to the complete \algo{}, we can observe that the recall and NDCG values by \algo{} w/o SMM are lower, particularly on larger datasets {\em Deezer-RO}, {\em Deezer-DBLP}, and {\em Youtube}. For instance, on a small dataset {\em Flickr}, the inclusion of SMM results in a margin of 0.0037 in Recall@5, while the improvements can be up to 0.0381 and 0.0375 on {\em Youtube}.
These results showcase the effectiveness of SMM and validate the necessity of incorporating global social structures in community recommendation.

\stitle{w/o SCA and NSC Choice in SCA}
As plotted in Fig. \ref{fig:ablation}, we can see that SCA is the most impactful component in \algo{} compared to others in the community recommendation performance. For example, without SCA, a significant drop of at least 0.0175 and 0.0375 in recall and NDCG can be observed on all datasets. On {\em Deezer-HR}, the performance decrease can be up to 0.1094 and 0.1185. These results manifest the vital role of local social structural features in community recommendation.  

Fig. \ref{fig:NSC} additionally shows the recall and NDCG results obtained by \algo{} equipped with various NSC metrics listed in Table~\ref{tab:metrics} in SCA.
We can observe that RAI takes the lead in most cases. The only exceptions are on the small datasets {\em BlogCatalog} and {\em Flickr}, where RAI is slightly inferior to CN, AAI, SI, and LHNI but still superior to baseline models in Table~\ref{tab:performance}.

\stitle{w/o UCE}
Similarly, our UCE module also improves the recommendation performance due to its power of rich collaborative features from known user-community memberships in $\C$.
On {\em Flickr}, {\em Deezer-HR}, and {\em Deezer-RO}, its effectiveness is more pronounced, which is only second to SCA.

\stitle{w/o FME}
As shown in Fig. \ref{fig:ablation}, FME also demonstrates notable efficacy in \algo{}. On small and medium-sized datasets, \algo{} is able to obtain a gain of 0.0024 and 0.0021 on average in recall and NDCG with FME, while the improvements on the large dataset {\em DBLP} are 0.0081 and 0.0080.

\stitle{w/o $\mathcal{L}_{\textnormal{KL}}$}
Another observation we can make is that the community detection loss $\mathcal{L}_{\textnormal{KL}}$ is indispensable for accurate community recommendation. More precisely, \algo{}'s performance will reduce by an average of 0.80\% and 1.95\% in Recall@5 and NDCG@5, respectively.
As pinpointed in Section~\ref{sec:training}, the benefit of $\mathcal{L}_{\textnormal{KL}}$ stems from its ability to guide the refinement of cluster centroids, i.e., community embeddings.

\begin{figure}[!t]
    \centering
    
    \begin{tikzpicture}
        \begin{customlegend}[
            legend entries={Recall@5, NDCG@5},
            legend columns=2,
            area legend,
            legend style={at={(0.5,1.05)},anchor=north,draw=none,font=\footnotesize,column sep=0.25cm}
        ]
            \addlegendimage{line legend, solid, thick, color=pythonBlue, mark=*, 
            mark options={solid, fill=pythonBlue, scale=1.2}}
            \addlegendimage{line legend, solid, thick, color=pythonOrange, mark=triangle*,
            mark options={solid, fill=pythonOrange, scale=1.2, rotate=0}}
        \end{customlegend}
    \end{tikzpicture}
    
    \vspace{-4mm}
    \subfloat[{\em BlogCatalog}]{%
        \includegraphics[width=0.32\columnwidth]{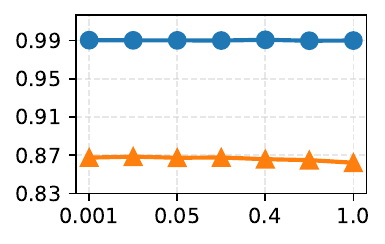}
    }
    \subfloat[{\em Flickr}]{%
        \includegraphics[width=0.31\columnwidth]{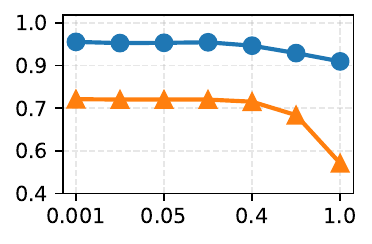}
    }
    \subfloat[{\em Deezer-HR}]{%
        \includegraphics[width=0.32\columnwidth]{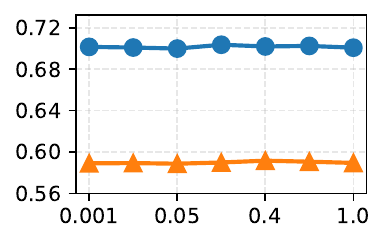} 
    }
    \vspace{-2ex}
    \subfloat[{\em Deezer-RO}]{%
        \includegraphics[width=0.32\columnwidth]{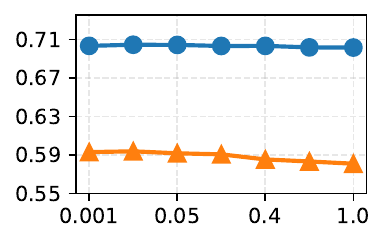} 
    }
    \subfloat[{\em DBLP}]{%
        \includegraphics[width=0.32\columnwidth]{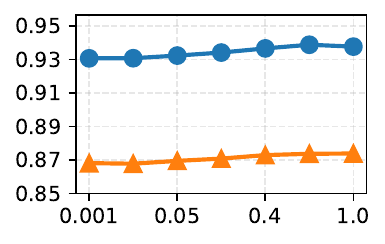}
    }
    \subfloat[{\em Youtube}]{%
        \includegraphics[width=0.32\columnwidth]{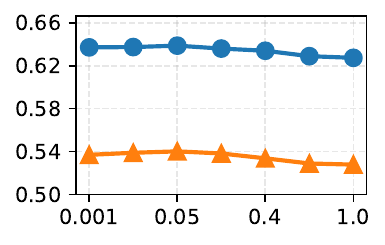}
    }
    \vspace{-1ex}
    \caption{Varying $\lambda$ in datasets.}
    \label{fig:vary-lambda}
\end{figure}

\subsection{Hyperparameter Study}\label{sec:exp-param}
Finally, we experimentally study the effects of hyperparameters $\beta$ in Eq.~\eqref{eq:emb} and $\lambda$ in Eq.~\eqref{eq:FME} on each dataset. Note that we fix $\alpha=0.33$ as it is required that $\alpha<\frac{1}{3}$ in Theorem~\ref{lem:SMM-sol}. Moreover, \algo{} is insensitive to $\gamma$ and $T$, which are set to $0.3$ and $2$ by default.

\stitle{Varying $\beta$}
Fig. \ref{fig:vary-beta} reports the Recall@5 and NDCG@5 results when varying $\beta$ from 0.1 to 1.0. It can be observed that as the value of $\beta$ increases, \algo{} exhibits diverse trends in performance on different datasets. 
Specifically, on {\em BlogCatalog} and {\em Flickr}, \algo{} remains stable, whereas its performance grows on {\em DBLP} and {\em Youtube} when $\beta$ is increased.
As for {\em Deezer-HR} and {\em Deezer-RO}, \algo{} attains a performance nadir when $\beta=0.3$, but a performance peak when $\beta$ is around $0.5$ or $0.6$.
Recall that $\beta$ is to balance the social features and collaborative features injected into the final user embeddings. Our results suggest that $0.5$ is a recommended setting for most datasets.

\stitle{Varying $\lambda$}
In Fig. \ref{fig:vary-lambda}, we display the recall that NDCG results when $\lambda$ is varied from 0.001 to 1.
As $\lambda$ increases, the performance of \algo{} remains stable on most datasets until $\lambda \le 0.05$, after which it experiences marginal decreases.
Particularly, on {\em Flickr}, a significant performance decline can be seen when $\lambda$ is increased from 0.4 to 1.0. This is due that the features in updated $\SM$ and $\BM$ are likely to be overwhelmed by the gradients from ${\SM}^{(i)}{\SM}^{(i)\top}$ and ${\BM}^{(i)}{\BM}^{(i)\top}$ as remarked in Section~\ref{sec:FME}.
Thus, the empirical results imply that a small $\lambda$ in the range of $[0.001, 0.05]$ is favorable and can mitigate such a feature's overwhelming issue.

\section{Conclusion}
This paper proposes an effective model \algo{} for social community recommendation. 
\algo{} includes four innovative and carefully-crafted components, SMM, SCA, UCE, and FME, which can fully and accurately extract and exploit social and collaborative signals inside the input data, for community recommendation.
Moreover, we additionally introduce a KL-based community detection loss into the model training of \algo{} so as to refine the community embeddings. 
Our extensive empirical evaluations demonstrate the superior performance of \algo{} to existing recommendation models in recommending communities on real-world social networks.
\begin{acks}
This work is supported by the National Natural Science Foundation of China (No. 62302414), the Hong Kong RGC ECS grant (No. 22202623), and the Huawei Gift Fund.
\end{acks}

\balance

\section*{GenAI Usage Disclosure}
We hereby disclose that no GenAI tools were used in code implementation, data analysis, manuscript writing, or any other aspects of the research.

\bibliographystyle{ACM-Reference-Format}
\bibliography{sample-base}

\newpage
\appendix
\section{Hilbert-Schmidt Independence Criterion}\label{sec:HSIC}

The Hilbert-Schmidt Independence Criterion (HSIC) \cite{gretton2005measuring_HSIC} is a statistical measure of dependency. Defined as the squared Hilbert-Schmidt norm, it quantifies the dependence between two variables by evaluating the cross-covariance operator within the Reproducing Kernel Hilbert Spaces. More specifically, for two random variables $\SM$ and $\BM$, the HSIC is rigorously defined as:
\begin{align*}
\texttt{HSIC}(\SM, \BM)
=& \left\| C_{\SM\BM} \right\|_{\text{HS}}^2 \\
=& \mathbb{E}_{\SM,\SM',\BM,\BM'} \left[ k_1(\SM, \SM') k_2(\BM, \BM') \right] \\
& + \mathbb{E}_{\SM,\SM'} \left[ k_1(\SM, \SM') \right] \mathbb{E}_{\BM,\BM'} \left[ k_2(\BM, \BM') \right] \\
& - 2 \mathbb{E}_{\SM,\BM} \left[ \mathbb{E}_{\SM'} \left[ k_1(\SM, \SM') \right] \mathbb{E}_{\BM'} \left[ k_2(\BM, \BM') \right] \right],
\end{align*}
where $\mathbb{E}$ denotes the expectation, and $k_1,k_2$ are two kernel functions for variables $\SM,\BM$. $\SM'$ and $\BM'$ are two independent copies of $\SM$ and $\BM$.

To develop a formal independence test grounded in the HSIC, we are required to approximate the HSIC value with a finite collection of observations. This approximation, referred to as the Empirical HSIC, is defined by the following formula:
\begin{equation}
\texttt{HSIC}(\SM, \BM) = \frac{1}{(|\U|-1)^{2}}\texttt{trace}(\textbf{K}^{(s)}\EM\textbf{K}^{(x)}\EM),
\end{equation}
where $|\U|$ is the number of independent observations. $\textbf{K}^{(s)}$ and $\textbf{K}^{(x)}$ are the Gram matrices with elements as $\textbf{K}_{i,j}^{(s)}=k_1(\SM_i,\SM_j)$ and $\textbf{K}_{i,j}^{(x)}=k_2(\BM_i,\BM_j)$, respectively. $\EM=\textbf{I}-\frac{1}{|\U|}\mathbf{1}$ is to center the Gram matrix to have zero mean.

To enhance computational efficiency, instead of multiplying the representations $\SM$ and $\BM$ by $\EM$, we can normalize them and adopt the linear kernel. Consequently, we derive the following simplified form~\cite{yang2022graph}:
\begin{equation}
\texttt{HSIC}(\SM, \BM) = \texttt{trace}(\SM^{\top}\BM\BM^{\top}\SM).
\end{equation}

\section{Theoretical Proofs}\label{sec:proof}
\begin{proof}[\bf Proof of Lemma~\ref{lem:modularity}]
In the community recommendation task, a user can join multiple communities. Therefore, we adjust $\delta(u_i,u_j)$ to be the number of communities that both users have joined. Then, we obtain $\delta(u_i,u_j) = \sum_{k=1}^{|\V|}\DeltaM_{i,k} \cdot \DeltaM_{j,k}$.

Next, denote the degree vector as $\dvec$, where $\dvec_i = \dvec_i$ for each index $i$, and $\sqrt{\dvec}$ is the element-wise square root of $\dvec$.
According to Eq.\eqref{eq:norm_modularity}, we can derive that
\begin{align*}
Q(\NAM) &= \sum_{k=1}^{|\V|}\sum_{(u_i,u_j)\in \EDG}\left( \NAM_{i,j} - \frac{\sqrt{\dvec}_i \cdot \sqrt{\dvec}_j}{|\EDG|}\right) \cdot \DeltaM_{i,k} \cdot \DeltaM_{j,k}\\
&= \sum_{k=1}^{|\V|} \DeltaM_{\cdot,k}^{\top}\left( \NAM - \frac{\sqrt{\dvec} \cdot \sqrt{\dvec}^{\top}}{|\EDG|}\right) \DeltaM_{\cdot,k} \\
&= \texttt{trace}\left(\DeltaM^{\top}\left(\NAM- \frac{\sqrt{\dvec}\sqrt{\dvec}^{\top}}{|\EDG|}\right)\DeltaM\right).
\end{align*}
\end{proof}

\begin{proof}[\bf Proof of Theorem~\ref{lem:SMM-sol}]
First, we need the following lemmata.
\begin{lemma}[\cite{horn2012matrix}]\label{col:neu}
Let $\MM$ be a matrix whose dominant eigenvalue $\lambda$ satisfies $|\lambda|<1$. Then, $\IM-\MM$ is invertible, and its inverse $(\IM-\MM)^{-1}$ can be expanded as a Neumann series: $(\IM-\MM)^{-1}=\sum_{t=0}^\infty\MM^t$.
\end{lemma}

\begin{lemma}\label{lem:range}
$\forall 1\le i\le |\U|$, $  -2\le  \lambda_i(\NAM- {\sqrt{\dvec}\sqrt{\dvec}^{\top}}/{|\EDG|}) \le 2$ where $\lambda_i(\cdot)$ returns the $i^{\textnormal{th}}$ eigenvalue.
\end{lemma}

Denote by $\mathcal{O}_{\textnormal{SMM}}$ Eq.~\eqref{eq:SMM-obj}. We take the derivative of $\mathcal{O}_{\textnormal{SMM}}$ w.r.t. $\PiM$ and set it to zero, which yields:
\begin{align}
& \frac{\partial \mathcal{O}_{\textnormal{SMM}}}{\partial \PiM}  = \alpha \cdot \left(\frac{\sqrt{\dvec}\sqrt{\dvec}^{\top}}{|\EDG|}-\NAM\right)\PiM + (1-\alpha)\cdot (\PiM-\UM^{\circ}) = 0 \notag\\
& \left( (1-\alpha)\IM - \alpha \left(\NAM - \frac{\sqrt{\dvec}\sqrt{\dvec}^{\top}}{|\EDG|}\right) \right) \PiM = (1-\alpha)\UM^{\circ} \notag \\
& \left( \IM - \frac{\alpha}{1-\alpha} \left(\NAM - \frac{\sqrt{\dvec}\sqrt{\dvec}^{\top}}{|\EDG|}\right) \right) \PiM = \UM^{\circ} \notag \\
& \PiM = \left( \IM - \frac{\alpha}{1-\alpha} \left(\NAM - \frac{\sqrt{\dvec}\sqrt{\dvec}^{\top}}{|\EDG|}\right) \right)^{-1}\UM^{\circ}\label{eq:SMM-closed-form-sol}.
\end{align}

When $\alpha < \frac{1}{3}$, we have $\frac{\alpha}{1-\alpha} < \frac{1}{2}$. Thus, the eigenvalues of $\frac{\alpha}{1-\alpha} (\NAM- \frac{\sqrt{\dvec}\sqrt{\dvec}^{\top}}{|\EDG|})$ are in the range of $(-1,1)$. Applying Lemma~\ref{col:neu} on Eq.~\eqref{eq:SMM-closed-form-sol} completes the proof.
\end{proof}

\begin{proof}[\bf Proof of Lemma~\ref{lem:range}]
First, we need to prove the eigenvalues of $\NAM$ and ${\sqrt{\dvec}\sqrt{\dvec}^{\top}}/{|\EDG|}$ are in the range of $[-1,1]$.

For $\NAM$, it is known that its dominant eigenvalue is bounded by $1$~\cite{chung1997spectral}. Thus, the range of its eigenvalues is within the interval $[-1,1]$.

For ${\sqrt{\dvec}\sqrt{\dvec}^{\top}}/{|\EDG|}$, we first prove that its Rayleigh quotient satisfies
\begin{equation*}
\frac{\xvec^{\top}\cdot \frac{\sqrt{\dvec}\sqrt{\dvec}^{\top}}{|\EDG|}\cdot \xvec}{\xvec^{\top}\xvec}\le 1.
\end{equation*}
First, we can derive that
\begin{align*}
\xvec^{\top} \left(\IM-\frac{\sqrt{\dvec}\sqrt{\dvec}^{\top}}{|\EDG|}\right) \xvec  =& \sum_{u_i\in \U}{\xvec_i^2} - \sum_{u_i,u_j\in \U}{\xvec_i\cdot \frac{\sqrt{\dvec_i\cdot \dvec_j}}{|\EDG|}\cdot \xvec_j}\\
=&  \sum_{u_i\in \U}{\xvec_i^2\cdot \frac{\dvec_i}{|\EDG|}} + \sum_{u_i\in \U}{\xvec_i^2\cdot \left(1-\frac{\dvec_i}{|\EDG|}\right)} \\
& - \sum_{u_i,u_j\in \U}{\xvec_i\cdot \sqrt{\frac{\dvec_i}{|\EDG|}}\cdot \xvec_j\cdot \sqrt{\frac{\dvec_j}{|\EDG|}}} \\
= & \frac{1}{2}\sum_{v_i\in \V}{\left(\xvec_i\cdot \sqrt{\frac{\dvec_i}{|\EDG|}}-\xvec_j\cdot \sqrt{\frac{\dvec_j}{|\EDG|}}\right)^2}\\
& + \sum_{u_i\in \U}{\xvec_i^2\cdot \left(1-\frac{\dvec_i}{|\EDG|}\right)}.
\end{align*}
Since $1-\frac{\dvec_i}{|\EDG|}\ge 0$, we have $\xvec^{\top} \left(\IM-\frac{\sqrt{\dvec}\sqrt{\dvec}^{\top}}{|\EDG|}\right) \xvec\ge 0$, meaning that $\frac{\xvec^{\top}\cdot \frac{\sqrt{\dvec}\sqrt{\dvec}^{\top}}{|\EDG|}\cdot \xvec}{\xvec^{\top}\xvec}\le 1$. This implies that the dominant eigenvalue of $\frac{\sqrt{\dvec}\sqrt{\dvec}^{\top}}{|\EDG|}$ is less than or equal to $1$. Its eigenvalues are in the range of $[-1,1]$.

Then, by Weyl inequality \cite{franklin2012matrix, weyl1912asymptotische} that for any Hermitian matrices $\AM$ and $\YM$, their dominant eigenvalues hold $\lambda_{i+j-1}(\AM + \YM) \leq \lambda_i(\AM) + \lambda_j(\YM) \leq \lambda_{i+j-n}(\AM + \YM)$ with $\lambda_1 \geq ... \geq \lambda_n$. Consequently, we can obtain 
\begin{align*}
\lambda_1 \left( \NAM - \frac{\sqrt{\dvec}\sqrt{\dvec}^{\top}}{|\EDG|}\right) 
&\leq \lambda_1(\NAM) + \lambda_1 \left( -\frac{\sqrt{\dvec}\sqrt{\dvec}^{\top}}{|\EDG|}\right) \\
&=  \lambda_1(\NAM) - \lambda_n \left(\frac{\sqrt{\dvec}\sqrt{\dvec}^{\top}}{|\EDG|}\right) \\
&\leq 1 - (-1) = 2,
\end{align*}
and
\begin{align*}
\lambda_n \left( \NAM - \frac{\sqrt{\dvec}\sqrt{\dvec}^{\top}}{|\EDG|}\right) 
&\geq \lambda_n(\NAM) + \lambda_n \left( -\frac{\sqrt{\dvec}\sqrt{\dvec}^{\top}}{|\EDG|}\right) \\
&=  \lambda_n(\NAM) - \lambda_1 \left(\frac{\sqrt{\dvec}\sqrt{\dvec}^{\top}}{|\EDG|}\right) \\
&\geq (-1) - 1 = -2.
\end{align*}
Based thereon, we derive $  -1\le  \lambda_i(\NAM- {\sqrt{\dvec}\sqrt{\dvec}^{\top}}/{|\EDG|}) \le 1$.
\end{proof}

\begin{proof}[\bf Proof of Theorem~\ref{lem:metrics}]
For Common Neighbors (CN), 
\begin{align*}
p(u_i, u_j) & = |\N_{\G}(u_i) \cap \N_{\G}(u_j)|\\
& = \sum_{k} \AM_{ik} \AM_{jk} = \AM_i \AM_j^\top = f(\AM)_i \cdot f(\AM)_j^\top .
\end{align*}
For Adamic-Adar Index (AAI),
\begin{align*}
p(u_i, u_j) &= \sum_{u \in \N(u_i)\cap \N(u_j)} \frac{1}{\log \dvec_u} = \sum_{u} \frac{\AM_{iu} \AM_{ju}}{\log \dvec_u}\\ &= \sum_{u} \frac{\AM_{iu}}{\sqrt{\log \dvec_u}}\cdot \frac{\AM_{ju}}{\sqrt{\log \dvec_u}}\\
& = \left(\AM\DM_{\log}^{-1/2}\right)_i \left(\AM\DM_{\log}^{-1/2}\right)^\top_j \\
&= f(\AM)_i \cdot f(\AM)_j^\top.
\end{align*}
For Resource Allocation Index (RAI),
\begin{align*}
p(u_i, u_j) &= \sum_{u \in \N_{\G}(u_i)\cap \N_{\G}(u_j)} \frac{1}{\dvec_u} = \sum_{u} \frac{\AM_{iu} \AM_{ju}}{\dvec_u}\\ &= \sum_{u} \frac{\AM_{iu}}{\sqrt{\log \dvec_u}} \cdot \frac{\AM_{ju}}{\sqrt{\log \dvec_u}}\\
& = \left(\AM\DM^{-1/2}\right)_i \left(\AM\DM^{-1/2}\right)^\top_j\\ 
&= f(\AM)_i \cdot f(\AM)_j^\top.
\end{align*}
For Salton Index (SI),
\begin{align*}
p(u_i, u_j) &= \frac{|\N_{\G}(u_i)\cap \N_{\G}(u_j)|}{\sqrt{\dvec_i \dvec_j}} = \sum_{u} \frac{\AM_{iu} \AM_{ju}}{\sqrt{\dvec_i \dvec_j}} \\
&= \sum_{u} \frac{\AM_{iu}}{\sqrt{\dvec_j}} \cdot \frac{\AM_{ju}}{\sqrt{\dvec_j}}\\
& = \left(\DM^{-1/2}\AM\right)_i \left(\DM^{-1/2}\AM\right)^\top_j \\
&=f(\AM)_i \cdot f(\AM)_j^\top.
\end{align*}
For Leicht-Holme-Newman Index (LHNI), 
\begin{align*}
p(u_i, u_j) &= \frac{|\N_{\G}(u_i)\cap \N_{\G}(u_j)|}{\dvec_i \dvec_j} = \sum_{u} \frac{\AM_{iu} \AM_{ju}}{\dvec_i \dvec_j} \\&= \sum_{u} \frac{\AM_{iu} \AM_{ju}}{\dvec_i \dvec_j} =\left(\DM^{-1}\AM\right) \left(\DM^{-1}\AM\right)^\top\\
&=f(\AM)_i \cdot f(\AM)_j^\top.
\end{align*}
The theorem is then proved.
\end{proof}

\begin{proof}[\bf Proof of Lemma~\ref{lem:B-bound}]
By the definition of $s(u_i,u_j)$ in Eq.~\eqref{eq:preference-sim},
\begin{align}
& \sum_{k=1}^{|\V|}\left(\hat{\YM}_{i,k} - \sqrt{\frac{\dtvec_i}{|\M|}} \cdot \sqrt{\frac{\svec_k}{|\M|}}\right)\cdot \left(\hat{\YM}_{j,k} - \sqrt{\frac{\dtvec_j}{|\M|}} \cdot \sqrt{\frac{\svec_k}{|\M|}}\right)\notag\\
= &\sum_{k=1}^{|\V|} \hat{\YM}_{i,k}\cdot \hat{\YM}_{j,k} - \hat{\YM}_{i,k}\cdot \sqrt{\frac{\dtvec_j}{|\M|}} \cdot \sqrt{\frac{\svec_k}{|\M|}}-\hat{\YM}_{j,k}\cdot \sqrt{\frac{\dtvec_i}{|\M|}} \cdot \sqrt{\frac{\svec_k}{|\M|}} \notag\\
& + \frac{\sqrt{\dtvec_i}\sqrt{\dtvec_j}\cdot \svec_k}{|\M|^2} \notag\\
= &\sum_{k=1}^{|\V|} \hat{\YM}_{i,k}\cdot \hat{\YM}_{j,k} + \sum_{k=1}^{|\V|}\frac{\sqrt{\dtvec_i}\sqrt{\dtvec_j}\cdot \svec_k}{|\M|^2} \notag\\
& - \sum_{k=1}^{|\V|} {\YM}_{i,k}\cdot {\frac{\sqrt{\dtvec_j}}{\sqrt{\dtvec_i}\cdot|\M|}} -\sum_{k=1}^{|\V|}{\YM}_{j,k}\cdot {\frac{\sqrt{\dtvec_i}}{\sqrt{\dtvec_j}\cdot|\M|}} \notag\\
& = \sum_{k=1}^{|\V|} \hat{\YM}_{i,k}\cdot \hat{\YM}_{j,k} + \frac{\sqrt{\dtvec_i}\sqrt{\dtvec_j}}{|\M|} - \frac{2 \sqrt{\dtvec_i}\sqrt{\dtvec_j}}{|\M|} \notag\\
& = \sum_{k=1}^{|\V|} \hat{\YM}_{i,k}\cdot \hat{\YM}_{j,k} - \frac{ \sqrt{\dtvec_i}\sqrt{\dtvec_j}}{|\M|}.\label{eq:s-bound}
\end{align}

Since $\YM_{i,k} \in \{0, 1\}$, we have
\begin{equation*}
\|\hat{\YM}_i\|_2^2=\sum_{k=1}^{|\V|}\hat{\YM}_{i,k}^2=\sum_{k=1}^{|\V|} \frac{\YM_{i,k}^2}{\dtvec_i \cdot \svec_k}=\sum_{k=1}^{|\V|} \frac{\YM_{i,k}}{\dtvec_i \cdot \svec_k}.
\end{equation*}
Here, we disregard the community with $\svec_k=0$, since they convey no information. Therefore, with $\svec_k\ge 1$,
\begin{equation*}
\|\hat{\YM}_i\|_2^2 = \sum_{k=1}^{|\V|} \frac{\YM_{i,k}}{\dtvec_i \cdot \svec_k} \leq \sum_{k=1}^{|\V|} \frac{\YM_{i,k}}{\dtvec_i} =1.
\end{equation*}

For any two vectors $\mathbf{a}, \mathbf{b} \in \mathbb{R}^n$, the Cauchy-Schwarz Inequality states
$|\mathbf{a} \cdot \mathbf{b}| \le \|\mathbf{a}\|_2 \cdot \|\mathbf{b}\|_2.$ By Cauchy-Schwarz, 
\begin{equation*}
\left|\sum_{k=1}^{|\V|} \hat{\YM}_{i,k} \cdot \hat{\YM}_{j,k}\right| \le \|\hat{\YM}_i\|_2 \cdot \|\hat{\YM}_j\|_2 \leq 1\cdot 1=1.
\end{equation*}

Since $\hat{\YM}_{i,k} \ge 0$, the dot product is non-negative
\begin{equation*}
\sum_{k=1}^{|\V|} \hat{\YM}_{i,k} \cdot \hat{\YM}_{j,k} \ge 0.    
\end{equation*}

Then, we can derive 
$0\le \sum_{k=1}^{|\V|} \hat{\YM}_{i,k}\cdot \hat{\YM}_{j,k} \le 1$, which completes the proof by plugging it into Eq.~\eqref{eq:s-bound}.
\end{proof}

\section{Experimental Details}\label{sec:exp-add}

\subsection{Datasets and Hyperparameter Settings}

We describe the details of each dataset used in the experiments in what follows:
\begin{itemize}[leftmargin=*]
\item {{\em BlogCatalog}~\cite{meng2019co-minidata}}: This is a network of social relationships of bloggers from the BlogCatalog website. In this scenario, we define the topic categories provided by the users as their preferred communities.
\item {\em Flickr~\cite{meng2019co-minidata}}: The data retrieved from a well-known site that serves as a hub for users to showcase their personal pictures and video snippets. The community is defined as the interest groups that are formed by users.
\item {{\em Deezer-HR} \& {\em Deezer-RO}~\cite{rozemberczki2019gemsec-Deezer}}:
The data are collected from the music streaming service Deezer. These datasets represent friendship networks of users from Croatia and Romania, respectively. In this dataset, we define the distinct music genres that users prefer as communities, which can be regarded as interest groups.
\item {\em DBLP~\cite{yang2012defining-largedata}}: 
The data is sourced from the DBLP computer science bibliography. Specifically, the social network under consideration is a co-authorship network where two authors are linked if they have co-authored at least one paper. Subsequently, the publication venue, such as a journal or a conference, serves as a ground-truth community. The community membership network represents authors who have published in a specific journal or conference.
\item {\em Youtube~\cite{yang2012defining-largedata}}: 
The data is collected from YouTube, a popular video-sharing website that incorporates a social network. In the YouTube social network, users can establish friendships with one another. Additionally, users have the ability to create groups to which other users can apply to join. For the purposes of this study, these user-defined groups are regarded as ground-truth communities.
\end{itemize}

We introduce the parameters that we did not mention in the main text. Some parameters are fixed for each dataset since it did not make a big difference for the experiment results, e.g., $\alpha$ in Eq.~\eqref{eq:SMM} is fixed as 0.33 for all datasets, the iteration rounds $T$ of Eq.~\eqref{eq:SMM} is fixed as 2, and $\gamma$ in Eq.~\eqref{eq:social-fusion} is fixed as 0.3. The hyperparameters $\lambda$ in Eq.~\eqref{eq:FME}, $\beta$ in Eq.~\eqref{eq:emb} and $\theta$ in Eq.~\eqref{loss:all} are the optimal parameters obtained through grid search on each dataset.

\begin{table}[!ht]
\small
\centering
\caption{Parameter setting in \algo{}}
\label{tab:params}
\vspace{-3ex}
\setlength{\tabcolsep}{3.5pt}  %
\begin{tabular}{@{}c*{6}{c}@{}}
\toprule
 & \multicolumn{6}{c}{Datasets} \\
\cmidrule(l{3pt}r{3pt}){2-7}
Parameter & {\em BlogCatalog} & {\em Flickr} & {\em Deezer-HR} & {\em Deezer-RO} & {\em DBLP} & {\em Youtube} \\
\midrule
$\alpha$ & 0.33 & 0.33 & 0.33 & 0.33 & 0.33 & 0.33 \\
$\gamma$ & 0.3  & 0.3 & 0.3 & 0.3 & 0.3 & 0.3 \\
$T$ & 2 & 2 & 2 & 2 & 2 & 2 \\
$\lambda$ & 0.01 & 0.01 & 0.1 & 0.05 & 0.7 & 0.1 \\
$\beta$   & 1 & 0.4 & 0.6 & 0.6 & 1 & 0.9 \\
$\theta$  & 1 & 1 & 0.05 & 0.5  & 0 & 0.01\\
\bottomrule
\end{tabular}
\end{table}

\subsection{Additional Experimental Results}

To comprehensively evaluate the algorithm's performance, we recorded Recall@$K$ and NDCG@$K$ results from top-1 to top-5 for each dataset. The best result is highlighted in \textbf{bold}, and the runner-up is \underline{underlined}.

The results clearly demonstrate that our proposed \algo{} almost consistently outperforms all baseline approaches across all tested datasets, with conspicuous improvements observed from top-1 to top-5 performance metrics. Specifically, on average across NDCG@1 to NDCG@5, \algo{} achieves considerable improvements over the best baselines, with gains of 2.3\%, 4.3\%, 3.2\%, 3.1\%, 5.6\%, and 6.4\% on the {\em BlogCatalog}, {\em Flickr}, {\em Deezer-HR}, and {\em Deezer-RO}, {\em DBLP} and {\em Youtube} respectively. The improvements achieved by \algo{} in recall are also significant, with recall gains typically ranging from 1\% to as high as 5\% across most datasets. For {\em Flickr} and {\em DBLP}, \algo{} demonstrates nearly identical performance to the state-of-the-art models \texttt{MHCN} and \texttt{GBSR} in Recall@5 while showing remarkable superiority across Recall@1 to Recall@5. It is worth noting that \algo often achieves substantial improvements in top-1 performance, with average gains of 5.9\% and 5.5\% on Recall@1 and NDCG@1, respectively. This indicates that by integrating global and local information, \algo{} can accurately identify the single community that users are most likely to be interested in.

On smaller datasets like {\em BlogCatalog} and {\em Flickr}, even optimal baseline methods such as \texttt{MHCN} integrate multiple triangle structures and self-supervised learning but show notable gaps in extracting global topological information. In contrast, our \algo{} consistently outperforms such baselines, with typical improvements ranging from 0.005 to 0.03 on {\em BlogCatalog}. Due to the data structure of {\em Flickr} possibly being more suited to triangular forms, our \algo{} shows slightly weaker performance than \texttt{MHCN} in Recall@5, while achieving average increases of 0.014 and 0.026 in other Recall and NDCG metrics, respectively. On medium-sized datasets like {\em Deezer-HR} and {\em Deezer-RO} (with dense community structures), our \algo{} stands out by effectively capturing close user-community connections. While \texttt{GBSR}, the denoise-based optimal baseline, removes interfering edges, it lacks the ability to model such community structures. As a result, \texttt{GBSR} averages a 0.02 decline relative to \algo{}. On large datasets {\em DBLP} and {\em Youtube}, which contain millions of social friendships, \algo{} achieves at least 1.0\% and 4.4\% improvements across Recall@1 to Recall@5 and NDCG@1 to NDCG@5 on {\em Youtube}. On {\em DBLP}, \algo{} achieves at least 0.3\% and 4.4\% improvements across Recall@1 to Recall@4 and NDCG@1 to NDCG@5, respectively, while \texttt{GBSR} occasionally shows better Recall@5 performance. The results highlight the efficacy of our \algo{} in leveraging and fusing social and collaborative information for community recommendation.

\begin{table*}[!t]
\centering
\renewcommand{\arraystretch}{0.9}
\caption{Additional experimental results on \em BlogCatalog.}
\vspace{-2ex}
\addtolength{\tabcolsep}{-0.2em}
\resizebox{\textwidth}{!}{%
    \begin{tabular}{c c c c c c c c c c c}
        \toprule
         \multirow{2}{*}{\bf Method} & \multicolumn{10}{c}{\em \bf BlogCatalog} \\
        \cmidrule(lr){2-6} \cmidrule(lr){7-11}  
        & Recall@1 & Recall@2 & Recall@3 & Recall@4 & Recall@5 & NDCG@1 & NDCG@2 & NDCG@3 & NDCG@4 & NDCG@5 \\
        \midrule
        \texttt{SVD++}~\cite{koren2008SVD++} & 0.1721 & 0.3297 & 0.5052 & 0.6686 & 0.8331 & 0.1721 & 0.2711 & 0.3587 & 0.4317 & 0.4854  \\
        \texttt{BPR}~\cite{rendle2009bpr} & 0.1628 & 0.3341 & 0.5069 & 0.6753 & 0.8387 & 0.1628 & 0.2720 & 0.3570 & 0.4282 & 0.4921 \\
        \texttt{LightGCN}~\cite{he2020lightgcn}& 0.1672 & 0.3378 & 0.5035 & 0.6713 & 0.8322 & 0.1672 & 0.2754 & 0.3590 & 0.4287 & 0.4920 \\
        \texttt{LightGCN-S}~\cite{yang2024GBSR} & 0.5762 & 0.8098 & 0.9026 & 0.9550 & 0.9831 & 0.5762 & 0.7254 & 0.7713 & 0.7935 & 0.8037 \\
        \texttt{DiffNet}~\cite{wu2019diffnet} & 0.6290 & 0.8165 & 0.8962 & 0.9534 & 0.9842 & 0.6290 & 0.7457 & 0.7875 & 0.8102 & 0.8232 \\
        \texttt{DiffNet++}~\cite{wu2020diffnet++} & 0.6195 & 0.8025 & 0.8856 & 0.9398 & 0.9747 & 0.6195 & 0.7329 & 0.7753 & 0.7991 & 0.8131\\
        \texttt{SEPT}~\cite{yu2021SEPT} & 0.1690 & 0.3489 & 0.5106 & 0.6836 & 0.8251 & 0.1690 & 0.2891 & 0.3664 & 0.4309 & 0.4991\\
        \texttt{MHCN}~\cite{yu2021MHCN} & \underline{0.6853} & \underline{0.8726} & \underline{0.9419} & \underline{0.9732} & \underline{0.9915} & \underline{0.6853} & \underline{0.8072} & \underline{0.8379} & \underline{0.8539} & \underline{0.8609}\\
        \texttt{GBSR}~\cite{yang2024GBSR} & 0.5822 & 0.7665 & 0.8530 & 0.9205 & 0.9646 & 0.5822 & 0.6976 & 0.7426 & 0.7712 & 0.7873\\
        \midrule
        \algo{} & \textbf{0.7127} & \textbf{0.8960} & \textbf{0.9517} & \textbf{0.9815} & \textbf{0.9962} & \textbf{0.7127} & \textbf{0.8271} & \textbf{0.8535} & \textbf{0.8676} & \textbf{0.8741}\\
        Improv. & \textbf{3.993\%} & \textbf{2.678\%} & \textbf{1.046\%} & \textbf{0.852\%} & \textbf{0.466\%} & \textbf{3.993\%} & \textbf{2.463\%} & \textbf{1.868\%} & \textbf{1.608\%} & \textbf{1.527\%} \\
        \bottomrule
    \end{tabular}
}
\vspace{2ex}
\end{table*}

\begin{table*}[!t]
\centering
\renewcommand{\arraystretch}{0.9}
\caption{Additional experimental results on \em Flickr.}
\vspace{-2ex}
\addtolength{\tabcolsep}{-0.2em}
\resizebox{\textwidth}{!}{%
    \begin{tabular}{c c c c c c c c c c c}
        \toprule
         \multirow{2}{*}{\bf Method} & \multicolumn{10}{c}{\em \bf Flickr} \\
        \cmidrule(lr){2-6} \cmidrule(lr){7-11}  
        & Recall@1 & Recall@2 & Recall@3 & Recall@4 & Recall@5 & NDCG@1 & NDCG@2 & NDCG@3 & NDCG@4 & NDCG@5 \\
        \midrule
        \texttt{SVD++}~\cite{koren2008SVD++} & 0.1086 & 0.2207 & 0.3373 & 0.4514 & 0.5599 & 0.1086 & 0.1813 & 0.2447 & 0.2839 & 0.3264\\
        \texttt{BPR}~\cite{rendle2009bpr} & 0.1077 & 0.2164 & 0.3358 & 0.4446 & 0.5612 & 0.1077 & 0.1810 & 0.2362 & 0.2858 & 0.3299\\
        \texttt{LightGCN}~\cite{he2020lightgcn} & 0.1090 & 0.2230 & 0.3352 & 0.4478 & 0.5566 & 0.1090 & 0.1806 & 0.2364 & 0.2851 & 0.3277\\
        \texttt{LightGCN-S}~\cite{yang2024GBSR}& 0.3679 & 0.5518 & 0.6705 & 0.7605 & 0.8352 & 0.3679 & 0.4841 & 0.5447 & 0.5824 & 0.6124 \\
        \texttt{DiffNet}~\cite{wu2019diffnet} & 0.4613 & 0.6371 & 0.7532 & 0.8306 & 0.8855 & 0.4613 & 0.5760 & 0.6309 & 0.6643 & 0.6838\\
        \texttt{DiffNet++}~\cite{wu2020diffnet++} & 0.3535 & 0.5322 & 0.6657 & 0.7605 & 0.8257 & 0.3535 & 0.4639 & 0.5310 & 0.5689 & 0.5974\\
        \texttt{SEPT}~\cite{yu2021SEPT} & 0.1684 & 0.2933 & 0.3950 & 0.4697 & 0.5741 & 0.1684 & 0.2492 & 0.3086 & 0.3478 & 0.3774\\
        \texttt{MHCN}~\cite{yu2021MHCN} & \underline{0.4882} & \underline{0.6921} & \underline{0.8069} & \underline{0.8916} & \textbf{0.9381} & \underline{0.4882} & \underline{0.6186} & \underline{0.6789} & \underline{0.7093} & \underline{0.7277}\\
        \texttt{GBSR}~\cite{yang2024GBSR} & 0.3226 & 0.4803 & 0.6004 & 0.6939 & 0.7752 & 0.3226 & 0.4221 & 0.4826 & 0.5227 & 0.5537\\
        \midrule
        \algo{} & \textbf{0.5241} & \textbf{0.7183} & \textbf{0.8191} & \textbf{0.8917} & \underline{0.9333} & \textbf{0.5281} & \textbf{0.6504} & \textbf{0.6996} & \textbf{0.7289} & \textbf{0.7466}\\
        Improv. & \textbf{7.355\%} & \textbf{3.776\%} & \textbf{1.522\%} & \textbf{0.015\%} & -0.507\% & \textbf{8.167\%} & \textbf{5.128\%} & \textbf{3.050\%} & \textbf{2.775\%} & \textbf{2.597\%} \\
        \bottomrule
    \end{tabular}
}
\vspace{2ex}
\end{table*}

\begin{table*}[!t]
\centering
\renewcommand{\arraystretch}{0.9}
\caption{Additional experimental results on \em Deezer-HR.}
\vspace{-2ex}
\addtolength{\tabcolsep}{-0.2em}
\resizebox{\textwidth}{!}{%
    \begin{tabular}{c c c c c c c c c c c}
        \toprule
         \multirow{2}{*}{\bf Method} & \multicolumn{10}{c}{\em \bf Deezer-HR} \\
        \cmidrule(lr){2-6} \cmidrule(lr){7-11}  
        & Recall@1 & Recall@2 & Recall@3 & Recall@4 & Recall@5 & NDCG@1 & NDCG@2 & NDCG@3 & NDCG@4 & NDCG@5 \\
        \midrule
        \texttt{SVD++}~\cite{koren2008SVD++} & 0.0732 & 0.1340 & 0.1864 & 0.2822 & 0.3187 & 0.1387 & 0.1518 & 0.1694 & 0.2017 & 0.2147\\
        \texttt{BPR}~\cite{rendle2009bpr} & 0.2899 & 0.4236 & 0.5213 & 0.5919 & 0.6467 & \underline{0.4921} & 0.4928 & 0.5169 & 0.5400 & 0.5605\\
        \texttt{LightGCN}~\cite{he2020lightgcn} & 0.2703 & 0.4084 & 0.5030 & 0.5703 & 0.6308 & 0.4814 & 0.4799 & 0.5018 & 0.5233 & 0.5462\\
        \texttt{LightGCN-S}~\cite{yang2024GBSR}& 0.2914 & 0.4313 & 0.5262 & 0.5952 & 0.6508 & 0.4895 & 0.4962 & 0.5194 & 0.5422 & 0.5638\\
        \texttt{DiffNet}~\cite{wu2019diffnet} & 0.1898 & 0.2801 & 0.3617 & 0.4176 & 0.4813 & 0.2913 & 0.3034 & 0.3351 & 0.3597 & 0.3832\\
        \texttt{DiffNet++}~\cite{wu2020diffnet++} & 0.1714 & 0.2757 & 0.3578 & 0.4149 & 0.4778 & 0.2620 & 0.2920 & 0.3240 & 0.3475 & 0.3713\\
        \texttt{SEPT}~\cite{yu2021SEPT} & 0.2670 & 0.4306 & 0.5331 & 0.6071 & 0.6576 & 0.4774 & 0.4964 & 0.5206 & 0.5442 & 0.5635\\
        \texttt{MHCN}~\cite{yu2021MHCN} & 0.0210 & 0.0804 & 0.1346 & 0.2085 & 0.2229 & 0.0540 & 0.0913 & 0.1144 & 0.1483 & 0.1532\\
        \texttt{GBSR}~\cite{yang2024GBSR} & \underline{0.2921} & \underline{0.4356} & \underline{0.5342} & \underline{0.6118} & \underline{0.6718} & 0.4884 & \underline{0.4982} & \underline{0.5234} & \underline{0.5503} & \underline{0.5738}\\
        \midrule
        \algo{} & \textbf{0.2946} & \textbf{0.4509} & \textbf{0.5602} & \textbf{0.6382} & \textbf{0.6983} & \textbf{0.4957} & \textbf{0.5159} & \textbf{0.5446} & \textbf{0.5723} & \textbf{0.5949}\\
        Improv. & \textbf{0.823\%} & \textbf{3.514\%} & \textbf{4.856\%} & \textbf{4.314\%} & \textbf{3.939\%} & \textbf{0.741\%} & \textbf{3.547\%} & \textbf{4.054\%} & \textbf{4.003\%} & \textbf{3.676\%} \\
        \bottomrule
    \end{tabular}
}
\vspace{2ex}
\end{table*}

\begin{table*}[!t]
\centering
\renewcommand{\arraystretch}{0.9}
\caption{Additional experimental results on \em Deezer-RO.}
\vspace{-2ex}
\addtolength{\tabcolsep}{-0.2em}
\resizebox{\textwidth}{!}{%
    \begin{tabular}{c c c c c c c c c c c}
        \toprule
         \multirow{2}{*}{\bf Method} & \multicolumn{10}{c}{\em \bf Deezer-RO} \\
        \cmidrule(lr){2-6} \cmidrule(lr){7-11}  
        & Recall@1 & Recall@2 & Recall@3 & Recall@4 & Recall@5 & NDCG@1 & NDCG@2 & NDCG@3 & NDCG@4 & NDCG@5 \\
        \midrule
        \texttt{SVD++}~\cite{koren2008SVD++} & 0.0000 & 0.0188 & 0.1127 & 0.1595 & 0.2174 & 0.0000 & 0.0160 & 0.0743 & 0.0958 & 0.1189\\
        \texttt{BPR}~\cite{rendle2009bpr} & \underline{0.2862} & 0.4194 & 0.5297 & 0.6090 & 0.6677 & \underline{0.4829} & 0.4850 & 0.5164 & 0.5444 & 0.5665\\
        \texttt{LightGCN}~\cite{he2020lightgcn} & 0.2783 & 0.4207 & 0.5152 & 0.5818 & 0.6371 & 0.4828 & 0.4884 & 0.5105 & 0.5312 & 0.5518\\
        \texttt{LightGCN-S}~\cite{yang2024GBSR}& 0.2816 & 0.4282 & 0.5337 & 0.6035 & 0.6609 & 0.4712 & 0.4862 & 0.5162 & 0.5402 & 0.5624\\
        \texttt{DiffNet}~\cite{wu2019diffnet}& 0.1765 & 0.2824 & 0.3569 & 0.4326 & 0.4964 & 0.2725 & 0.3005 & 0.3263 & 0.3550 & 0.3840\\
        \texttt{DiffNet++}~\cite{wu2020diffnet++}& 0.1719 & 0.2688 & 0.3511 & 0.4246 & 0.4985 & 0.2646 & 0.2863 & 0.3187 & 0.3491 & 0.3800\\
        \texttt{SEPT}~\cite{yu2021SEPT} & 0.2760 & \underline{0.4332} & \underline{0.5357} & 0.6093 & 0.6644 & 0.4801 & \underline{0.4979} & \underline{0.5243} & \underline{0.5482} & \underline{0.5688}\\
        \texttt{MHCN}~\cite{yu2021MHCN} & 0.2714 & 0.4214 & 0.5351 & 0.6057 & 0.6628 & 0.4707 & 0.4842 & 0.5165 & 0.5396 & 0.5612\\
        \texttt{GBSR}~\cite{yang2024GBSR}& 0.2824 & 0.4313 & 0.5323 & \underline{0.6111} & \underline{0.6750} & 0.4705 & 0.4880 & 0.5160 & 0.5433 & 0.5683\\
        \midrule
        \algo{} & \textbf{0.2969} & \textbf{0.4578} & \textbf{0.5643} & \textbf{0.6405} & \textbf{0.7019} & \textbf{0.4915} & \textbf{0.5128} & \textbf{0.5423} & \textbf{0.5683} & \textbf{0.5906}\\
        Improv. & \textbf{3.729\%} & \textbf{5.685\%} & \textbf{5.335\%} & \textbf{4.809\%} & \textbf{3.984\%} & \textbf{1.779\%} & \textbf{2.992\%} & \textbf{3.434\%} & \textbf{3.659\%} & \textbf{3.837\%} \\
        \bottomrule
    \end{tabular}
}
\vspace{2ex}
\end{table*}

\begin{table*}[!t]
\centering
\renewcommand{\arraystretch}{0.9}
\caption{Additional experimental results on \em DBLP.}
\vspace{-2ex}
\addtolength{\tabcolsep}{-0.2em}
\resizebox{\textwidth}{!}{%
    \begin{tabular}{c c c c c c c c c c c}
        \toprule
         \multirow{2}{*}{\bf Method} & \multicolumn{10}{c}{\em \bf DBLP} \\
        \cmidrule(lr){2-6} \cmidrule(lr){7-11}  
        & Recall@1 & Recall@2 & Recall@3 & Recall@4 & Recall@5 & NDCG@1 & NDCG@2 & NDCG@3 & NDCG@4 & NDCG@5 \\
        \midrule
        \texttt{SVD++}~\cite{koren2008SVD++} & 0.0001 & 0.0002 & 0.0006 & 0.0028 & 0.0087 & 0.0001 & 0.0002 & 0.0004 & 0.0014 & 0.0036\\
        \texttt{BPR}~\cite{rendle2009bpr} & 0.0688 & 0.1218 & 0.1666 & 0.2055 & 0.2339 & 0.0713 & 0.1031 & 0.1256 & 0.1416 & 0.1532\\
        \texttt{LightGCN}~\cite{he2020lightgcn} & 0.0948 & 0.1343 & 0.1572 & 0.1731 & 0.1887 & 0.1004 & 0.1218 & 0.1332 & 0.1402 & 0.1447\\
        \texttt{LightGCN-S}~\cite{yang2024GBSR}& 0.6922 & 0.8446 & 0.8993 & 0.9230 & 0.9366 & 0.7082 & 0.7942 & 0.8215 & 0.8323 & 0.8372\\
        \texttt{DiffNet}~\cite{wu2019diffnet}& 0.5387 & 0.6632 & 0.7326 & 0.7704 & 0.7993 & 0.5478 & 0.6206 & 0.6556 & 0.6728 & 0.6836\\
        \texttt{DiffNet++}~\cite{wu2020diffnet++}& 0.6884 & 0.7783 & 0.8130 & 0.8296 & 0.8436 & 0.6978 & 0.7458 & 0.7633 & 0.7707 & 0.7763\\
        \texttt{SEPT}~\cite{yu2021SEPT} & 0.2600 & 0.3467 & 0.3925 & 0.4136 & 0.4231 & 0.2655 & 0.3137 & 0.3386 & 0.3474 & 0.3514\\
        \texttt{MHCN}~\cite{yu2021MHCN} & \underline{0.7220} & 0.8216 & 0.8663 & 0.8885 & 0.9037 & \underline{0.7374} & 0.7840 & 0.8024 & 0.8121 & 0.8170\\
        \texttt{GBSR}~\cite{yang2024GBSR}& 0.6948 & \underline{0.8491} & \underline{0.9039} & \underline{0.9285} & \textbf{0.9424} & 0.7113 & \underline{0.7977} & \underline{0.8251} & \underline{0.8350} & \underline{0.8398}\\
        \midrule
        \algo{} & \textbf{0.7791} & \textbf{0.8814} & \textbf{0.9146} & \textbf{0.9317} & \underline{0.9415} & \textbf{0.7947} & \textbf{0.8493} & \textbf{0.8656} & \textbf{0.8730} & \textbf{0.8768}\\
        Improv. & \textbf{7.913\%} & \textbf{3.800\%} & \textbf{1.182\%} & \textbf{0.340\%} & -0.099\% & \textbf{7.761\%} & \textbf{6.460\%} & \textbf{4.904\%} & \textbf{4.555\%} & \textbf{4.404\%} \\
        \bottomrule
    \end{tabular}
}
\vspace{2ex}
\end{table*}

\begin{table*}[!t]
\centering
\renewcommand{\arraystretch}{0.9}
\caption{Additional experimental results on \em Youtube.}
\vspace{-2ex}
\addtolength{\tabcolsep}{-0.2em}
\resizebox{\textwidth}{!}{%
    \begin{tabular}{c c c c c c c c c c c}
        \toprule
         \multirow{2}{*}{\bf Method} & \multicolumn{10}{c}{\em \bf Youtube} \\
        \cmidrule(lr){2-6} \cmidrule(lr){7-11}  
        & Recall@1 & Recall@2 & Recall@3 & Recall@4 & Recall@5 & NDCG@1 & NDCG@2 & NDCG@3 & NDCG@4 & NDCG@5 \\
        \midrule
        \texttt{SVD++}~\cite{koren2008SVD++} & 0.0001 & 0.0003 & 0.0003 & 0.0005 & 0.0006 & 0.0001 & 0.0002 & 0.0003 & 0.0003 & 0.0004\\
        \texttt{BPR}~\cite{rendle2009bpr}& 0.0518 & 0.0828 & 0.1059 & 0.1256 & 0.1414 & 0.0679 & 0.0795 & 0.0903 & 0.0976 & 0.1048\\
        \texttt{LightGCN}~\cite{he2020lightgcn} & 0.1068 & 0.1589 & 0.1908 & 0.2159 & 0.2346 & 0.1355 & 0.1541 & 0.1680 & 0.1783 & 0.1861\\
        \texttt{LightGCN-S}~\cite{yang2024GBSR}& 0.3262 & 0.4436 & 0.5122 & 0.5581 & 0.5929 & 0.3624 & 0.4181 & 0.4501 & 0.4697 & 0.4838\\
        \texttt{DiffNet}~\cite{wu2019diffnet}& 0.1818 & 0.2579 & 0.3097 & 0.3479 & 0.3775 & 0.1924 & 0.2359 & 0.2618 & 0.2788 & 0.2907\\
        \texttt{DiffNet++}~\cite{wu2020diffnet++}& 0.2791 & 0.3460 & 0.3852 & 0.4164 & 0.4445 & 0.2875 & 0.3252 & 0.3435 & 0.3588 & 0.3692\\
        \texttt{SEPT}~\cite{yu2021SEPT} & 0.1224 & 0.1762 & 0.2133 & 0.2386 & 0.2603 & 0.1483 & 0.1693 & 0.1858 & 0.1969 & 0.2055\\
        \texttt{MHCN}~\cite{yu2021MHCN} & 0.2246 & 0.3126 & 0.3709 & 0.4156 & 0.4538 & 0.2610 & 0.2970 & 0.3237 & 0.3415 & 0.3543\\
        \texttt{GBSR}~\cite{yang2024GBSR}& \underline{0.3541} & \underline{0.4815} & \underline{0.5540} & \underline{0.6000} & \underline{0.6348} & \underline{0.3943} & \underline{0.4535} & \underline{0.4870} & \underline{0.5067} & \underline{0.5203}\\
        \midrule
        \algo{} & \textbf{0.3946} & \textbf{0.5066} & \textbf{0.5681} & \textbf{0.6107} & \textbf{0.6413} & \textbf{0.4350} & \textbf{0.4849} & \textbf{0.5132} & \textbf{0.5316} & \textbf{0.5435}\\
        Improv. & \textbf{11.447\%} & \textbf{5.208\%} & \textbf{2.540\%} & \textbf{1.778\%} & \textbf{1.038\%} & \textbf{10.335\%} & \textbf{6.930\%} & \textbf{5.380\%} & \textbf{4.907\%} & \textbf{4.443\%} \\
        \bottomrule
    \end{tabular}
}
\vspace{2ex}
\end{table*}

\end{document}